\pdfoutput=1
\documentclass[%
reprint,
amsmath,amssymb,
nofootinbib,
aps,
floatfix,
]{revtex4-1}
\usepackage{graphicx}
\usepackage{dcolumn}
\usepackage{afterpage}
\usepackage{textcomp}
\usepackage{gensymb}
\usepackage{placeins}
\usepackage{bm}
\usepackage{amsmath,amssymb}
\usepackage{subcaption}
\usepackage{float}
\usepackage{soul}
\usepackage{array}
\usepackage{epstopdf}
\usepackage{lipsum}
\usepackage[dvipsnames,svgnames,x11names]{xcolor}
\usepackage[final]{changes}

\begin{document}

\preprint{APS/123-QED}

\title{A primer on the oxDNA model of DNA: When to use it, how to simulate it and how to interpret the results}%

\author{A. Sengar}
 \altaffiliation[]{a.sengar@imperial.ac.uk}

\author{T. E. Ouldridge}
 \email{t.ouldrdige@imperial.ac.uk}
\affiliation{Department of Bioengineering and Centre for Synthetic Biology, Imperial College London, UK}

\author{O. Henrich}
\email{oliver.henrich@strath.ac.uk}
\affiliation{Department of Physics, SUPA, University of Strathclyde, Glasgow, UK}

\author{L. Rovigatti}
\email{lorenzo.rovigatti@uniroma1.it}
\affiliation{Department of Physics, Sapienza University of Rome, Italy}
\affiliation{CNR Institute of Complex Systems, Uos Sapienza, Italy}

\author{P. \v{S}ulc}
\email{psulc@asu.edu}
\affiliation{Center for Molecular Design and Biomimetics, The Biodesign Institute and School of Molecular Sciences, Arizona State University, USA}

\date{\today}

\begin{abstract}
The oxDNA model of DNA has been applied widely to systems in biology, biophysics and nanotechnology. It is currently available via two independent open source packages. Here we present a set of clearly-documented exemplar simulations that simultaneously provide both an introduction to simulating the model, and a review of the model's fundamental properties. We outline how simulation results can be interpreted in terms of -- and feed into our understanding of -- less detailed models that operate at larger length scales, and provide guidance on whether simulating a system with oxDNA is worthwhile.

\end{abstract}

\pacs{Valid PACS appear here}
\maketitle

\section{\label{sec:level1}Introduction}

Deoxyribonucleic acid (DNA) is a macromolecule that acts as a storage medium for genetic information for all living organisms \cite{Alberts2002}. In nature, the molecule is most often found as a double helix of two strands. The structure of each strand comprises of a backbone of covalently linked sugar and phosphate groups. Each sugar is further attached to a base moiety: adenine (A), guanine (G), cytosine (C) or thymine (T). Certain intra- and intermolecular interactions between these bases drive the formation of the aforementioned mentioned double helical structure. 

Crucially, the base pairing that holds these duplexes together is highly specific; to a first approximation, A will only bind to T and C will only bind to G, and vice versa. Matching -- or complementary -- sequences therefore bind to each other much more strongly than to non-complementary sequences. The different base identities, along with the rules of complementarity, allow information to be encoded into the single strands and copied from generation to generation \cite{Watson1953MolecularAcid}.

The DNA double helix has a diameter of about 2\,nm, and a helical pitch of about 3.4–3.6\,nm. Double strands are relatively stiff, with large bending disfavoured on lengthscales below around 40-50\,nm \cite{Seeman2003DNAWorld}. By contrast, single strands are very flexible \cite{Murphy2004ProbingSpectroscopy,Chen2012IonicDNA} forming loops and kinks with only a handful of bases or fewer. 

These thermodynamic, mechanical and structural properties influence DNA's biological role, but also make it an ideal material for nanoscale engineering. The simplicity of interactions between strands, and the predictability of the structural and mechanical properties of the product, 
have enabled the rational design of a host of synthetic structures \cite{Fu93,Goodman2005,rothemund2006folding,Douglas09,Ke2012,zhang2015complex,tikhomirov2017fractal,wagenbauer2017gigadalton}, computing architectures \cite{Adleman1994MolecularProblems,rothemund2004algorithmic,qian2011neural,cherry2018scaling,woods2019diverse} and dynamic systems \cite{yurke2000dna,shin2004synthetic,Muscat2011,Wickham2012,zhang2011dynamic,tomov2017dna,srinivas2017enzyme}.

DNA's importance to biology, nanotechnology and simply as a canonical model biopolymer for biophysicists means that modelling its behaviour is a key challenge. Unsurprisingly, therefore, models spanning an enormous range of complexity have been proposed to analyse and rationalize the behaviour of DNA. In this pedagogical review, we will first discuss this range of models and their interplay, before focusing on a particular coarse-grained model, oxDNA. 

The oxDNA model, first published in 2010 \cite{Ouldridge2010DNADNA} (and with a slightly updated potential in 2011 \cite{Ouldridge2011StructuralModel}), has now been extensively applied to problems in nanotechnology \cite{Doye2013Coarse-grainingNanotechnology,Snodin2016DirectOrigami,Snodin2019Coarse-grainedOrigami,Ouldridge2013OptimizingWalker,Srinivas2013,Machinek2014ProgrammableDisplacement,Henning-Knechtel2017NARChannels,Hong2018Layered-CrossoverEngineering}, soft matter \cite{Stoev2020OnHydrogels,DeMichele2012Self-assemblyLink,Procyk2020Coarse-grainedNanotechnology,Rovigatti2014AccurateNanostars}, biophysics \cite{Nomidis2019Twist-bendBeyond,Romano2013Coarse-grainedOverstretching,Matek2012,Matek2015,harrison2019identifyingmodel,mosayebi2015force} and biology \cite{Wang2015Twist-inducedDensity,Lee2015BaseRecombinases,Craggs2019SubstratePolymerase}. Numerous tools exist to generate and visualize systems with oxDNA \cite{Henrich2018Coarse-grainedApplications,Suma2019TacoxDNA:Origami}, alongside two independent, publicly-available code bases for actually running simulations with at least three qualitatively distinct algorithms for simulating the model \cite{Snodin2015IntroducingDNA,Ouldridge2011StructuralModel}. One of these code bases has recently been incorporated into a webserver \cite{Poppleton2020DesignSimulation}. 

Despite this uptake, however, there is insufficient clarity on how the basic properties of the oxDNA model make it well- or poorly-suited to studying certain systems. Moreover, many interesting phenomena require non-trivial simulation techniques if they are to be probed with oxDNA. Although those techniques have been widely applied, and software implementing them with oxDNA is available, documentation supporting their use is limited. Equally, there is very little help with the intuition required to use these techniques successfully. Finally, a major aspect to interpreting the results from oxDNA is rationalizing its predictions in terms of less detailed models. Unfortunately, however, there are many subtleties in doing so.

In this pedagogical review we implement a series of exemplar simulations that allow us to address these shortcomings. These simulations will establish a well-documented set of examples for a series of approaches that can be adapted by users, and this review will provide some of the intuition for how to use these approaches successfully. Simultaneously, we will use these examples to illustrate key aspects of the oxDNA model that determine its usefulness, and will explore how to interpret the results in terms of DNA models at different scales. 

\section{DNA models across length scales}
\label{sec:models}
At the smallest and most fundamental scale, quantum chemistry calculations can be used to estimate the nucleotide properties from first principles \cite{Sponer2004AccuratePairs,Perez2004TheAnalysis,Hobza1999StructureCalculations,Sponer2008NatureBases}. However, these calculations are computationally extremely expensive and are unable to capture the collective behaviour of whole strands in solution. Nonetheless, insight from this field has been incorporated into classical atomistic force fields AMBER \cite{Cornell199651795197}  and CHARMM \cite{Brooks1983CHARMM:Calculations} that use empirical force fields to model interactions between atoms. These force fields are iteratively parameterised using both comparison to experimental data and  information from lower-level quantum mechanical descriptions. In recent years, advances in computational resources have allowed these models to simulate large systems -- such as DNA origami -- for long enough timescales to analyse their equilibrium properties. Given long simulations, these atomistic models are able to sample the conformation of large structures \cite{nguyen2014folding,rocklin2017global} and the breaking and formation of base pairs \cite{brown2015stacking}. However, at the time of writing, a systematic study of DNA duplex formation thermodynamics, as represented by atomistic models, has not been performed. As such, it is unknown how well these atomistic models represent DNA thermodynamics -- historically, the force fields have required adjustment as new systems and longer time scales are studied \cite{yoo2012improved,perez2007refinement}. This fact, alongside the heavy computational load in simulating large systems or significant structural changes, mean that atomistic approaches are currently limited to a fraction of the systems of interest in DNA-based biophysics, biology, soft matter and nanotechnology. 

In an effort to access longer timescales, a number of ``coarse-grained" or ``mesoscale" models have been introduced \cite{dans2016multiscale,hinckley2013experimentally,Savelyev2009MolecularDNA, ivani2016parmbsc1,machado2015exploring, korolev2014coarse,uusitalo2015martini,Ouldridge2011StructuralModel,maffeo2014coarse,maffeo2020mrdna,maciejczyk2014dna}. These models represent DNA with a much-reduced set of degrees of freedom relative to atomistic approaches. In particular, solvent (and solvated ions) are usually treated implicitly, and groups of atoms in the DNA are replaced by a single site with effective interactions. As a result, these models can access longer length and time scales than atomistic descriptions.

The procedure for coarse-graining ranges from ``bottom-up" approaches that seek to formally map the statistical behaviour of a more detailed model into a coarse-grained description \cite{maffeo2014coarse,maciejczyk2014dna,Savelyev2009MolecularDNA}, to ``top-down" approaches such as oxDNA that are more {\it ad hoc}, instead seeking to reproduce as many experimentally relevant properties as possible \cite{Ouldridge2009TheModel, maffeo2020mrdna,machado2015exploring,uusitalo2015martini,hinckley2013experimentally}. Bottom-up approaches have been most successfully used to study fluctuations within the duplex state, where the atomistic models on which they are built are best parameterised. Top-down approaches, by contrast, have found their application in the analysis of processes that involve DNA outside of its canonical B-form, including duplex hybridization \cite{Ouldridge2013DNADependence}, strand displacement \cite{Srinivas2013,irmisch2020modeling}, stress-induced structural transitions \cite{Romano2013Coarse-grainedOverstretching,wang2014modeling,Sutthibutpong2016Long-rangeSimulation} and the properties of nanostructures with branched helices and single-stranded sections \cite{Rovigatti2014AccurateNanostars,engel2020measuring}.

Although highly-simplified, all of the coarse-grained models cited above attempt to represent the discrete, three-dimensional structure of DNA explicitly. An important role in our understanding of DNA is played by even simpler models. In thermodynamic terms, two classes of model have received particular attention. Firstly, the Peyrard-Bishop-Dauxois model and its variants have been used to probe the statistical  properties of the duplex denaturation transition in the thermodynamic limit \cite{Dauxois1993DynamicsDenaturation,Nisoli2011ThermomechanicsLoad,Cocco1999StatisticalDNA}. These models represent DNA through two or three continuous degrees of freedom per base pair.

A second approach dispenses with continuous degrees of freedom altogether, taking an Ising-like approach in which base pairs are either present or absent. Originally introduced by Poland and Scheraga to probe the duplex denaturation phase transition  \cite{Poland1966OccurrenceModels}, the approach was adapted and carefully parameterized \cite{SantaLucia1998AThermodynamics,SantaLucia2004TheMotifs,Huguet2010,bae2020high} to describe binding equilibria for strands of moderate length (oligonucleotides). It is difficult to overstate just how influential the nearest neighbour model has been, particularly in the development of nucleic acid nanotechnology, as it allows rational design of an ensemble of strands to produce the desired thermodynamics. The NUPACK software suite automates this process of system analysis and thermodynamics-based design by implementing the nearest-neighbour model \cite{SantaLucia2004TheMotifs}. A number of attempts have been made to augment this thermodynamic model with realistic kinetics \cite{Srinivas2013,flamm2000rna,xayaphoummine2005kinefold,schaeffer2015stochastic}.

At its simplest, the nearest-neighbour model allows a two-state approximation to the binding of $A$ and $B$, in which the strands are either fully bound or fully dissociated. In this limit, the concentration of the product $[AB]$ can be estimated using the equation
 \begin{equation}
\dfrac{[AB]}{[A][B]}=\exp(- (\Delta H_{AB}-T\Delta S_{AB})/kT).     
 \end{equation}
 Here $\Delta H_{AB}$ and $\Delta S_{AB}$ are computed by summing contributions from each nearest-neighbor set of two base pairs, together with terms for helix initiation and various structural features, all of which are assumed to be temperature independent.  
 
Another class of models ignores thermodynamics entirely, instead providing a continuum-level description of DNA mechanics. Most notably, DNA is frequently modelled as a semi-flexible polymer (or worm-like chain, WLC) characterised by a bending modulus \cite{kratky1949rontgenuntersuchung}. This model  can be augmented with an extensional modulus \cite{Odijk1995} and a representation of twist with associated twist modulus \cite{yamakawa1977hypothesis}. It is also possible to consider coupling between the modes of deformation \cite{Nomidis2019Twist-bendBeyond,gore2006dna}. As with the nearest-neighbour model of DNA, the influence of these approaches is enormous, particularly within the biophysics community. The elastic rod is the starting point for understanding the geometry of DNA, and the null model against which results are compared and interpreted.   

There is actually quite a large gap in complexity between mesoscopic models such as oxDNA and the continuum WLC models or the nearest neighbour model of thermodynamics. The time required to analyse the same system with these methods differs by many orders of magnitude. It is intriguing that, to our knowledge, there are few approaches that come close to bridging this gap. Fundamentally, it is not easy to combine the mechanics of semiflexible DNA as captured by the WLC, the geometry and topology f DNA structures, and the thermodynamics of DNA duplex formation as described in the nearest-neighbour model, in a representation that is simultaneously quantitatively useful and substantially simpler than the existing mesoscale models. Approaches such as Benham's description of melting in circularly negatively-supercoiled DNA \cite{FyeExact1999} achieve this marriage in specific contexts. The variety of possible behaviour, however, and the sensitive interplay of topology, structure, mechanics and thermodynamics in many systems of interest, make the development of such models extremely hard and currently necessitate the application of coarse-grained models such as oxDNA.  

In the rest of this pedagogical review, we first provide a high-level description of the basics of the oxDNA model and simulation techniques. We then present prototypical simulations to demonstrate key properties of oxDNA, and discuss how results from these simulations can be interpreted in terms of simpler DNA models at different length scales. While doing so, we discuss specific challenges in obtaining meaningful data from oxDNA simulations, and discuss where oxDNA provides added value. Initialisation files, processing scripts and supporting instructions are provided for all simulations presented here at \cite{supportfiles}. This review should then serve as an introductory tutorial to applying oxDNA. 

One drawback of this format is that the examples are presented as a fait accompli; just re-running the code will provide a limited experience of the real process of simulating oxDNA. We strongly encourage readers using this document as a tutorial to attempt to construct as much as possible of the simulations for themselves, and then to compare to the results obtained here. Alternatively, users may try to construct variants to simulate similar systems. Additional guidance on the nuts and bolts of running simulations can be found at Ref. \cite{oxDNAwebsite,oxDNALAMMPS}, where instructions on visualizing the output can also be found. In general, we have found that checking one or two snapshots of a simulation can avoid many wasted hours simulating and studying faulty systems. 

\begin{figure*}
    \centering
            \includegraphics[width=\textwidth]{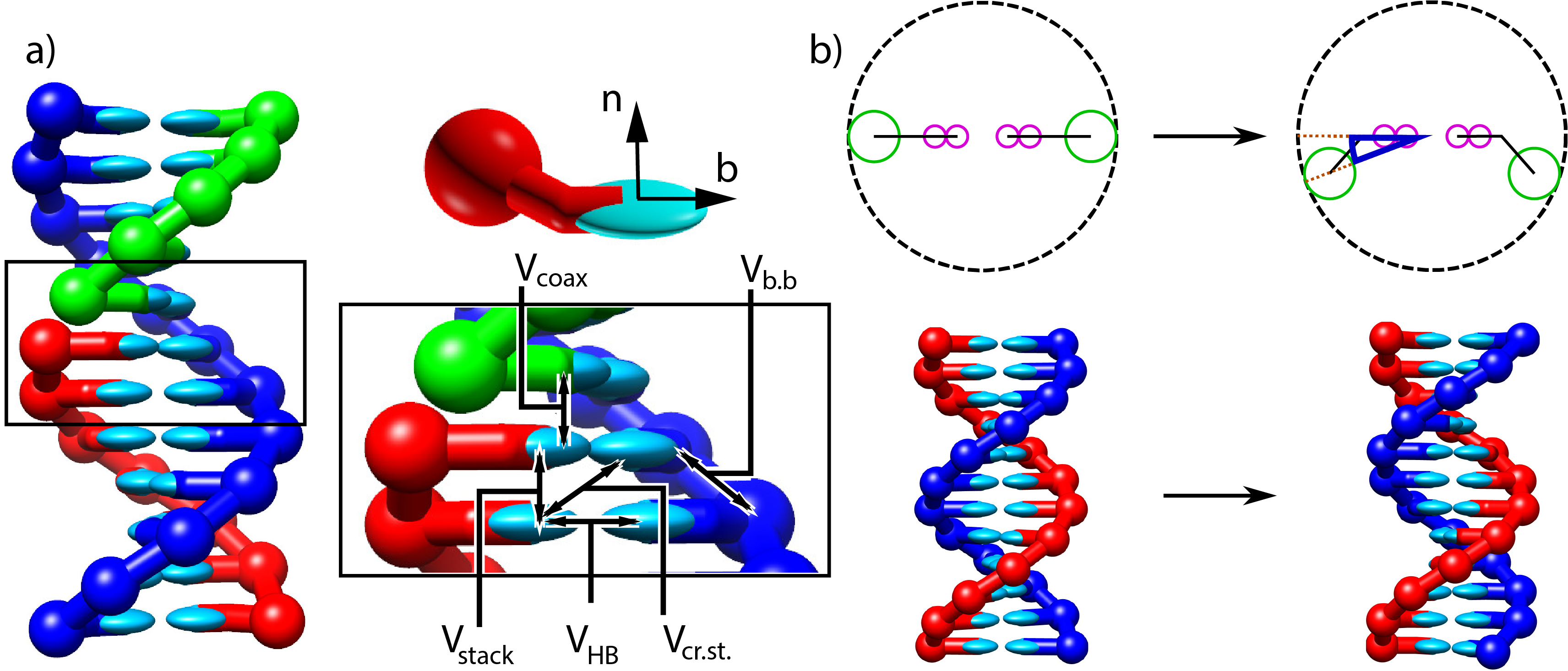}
        \caption
        {\small Structure and interactions of the oxDNA model (adapted from \cite{Snodin2015IntroducingDNA,DoyeOrigami2020}). a) Three strands forming a nicked duplex as represented by oxDNA2.0, with the central section of the complex illustrating key interactions from Eq.~\ref{eq:potential} highlighted. Individual nucleotides have an orientation described by a vector normal to the plane of the base (labelled n), and a vector indicating the direction of the hydrogen bonding interface (labelled b). b) Comparison of structure in oxDNA1.0 an oxDNA1.5 vs oxDNA2.0.  In the earlier version of the model, all interaction sites are co-linear; in oxDNA2.0, offsetting the backbone site allows for major and minor grooving. } 
   \label{fig:interaction_sites}
\end{figure*}

\section{The oxDNA model}
The oxDNA model was originally developed to study the self-assembly, structure and mechanical properties of DNA nanostructures, and the action of DNA nanodevices - although it has since been applied more broadly. To describe such systems, a model needs to capture the structural, mechanical and thermodynamic properties of single-stranded DNA, double-stranded DNA, and the transition between the two states. It must also be feasible to simulate large enough systems for long enough to sample the key phenomena. As discussed in Section~\ref{sec:models}, mesoscopic models in which multiple atoms are represented by a single interaction site are the appropriate resolution for these goals. 

We will now outline the key features of the oxDNA model, the specific mesoscale model that is the focus of this review. While doing so, we note that there are effectively three versions of the oxDNA potential that are publicly available. The original model, oxDNA1.0 \cite{Ouldridge2011StructuralModel}, lacks sequence-specific interaction strengths, electrostatic effects and major/minor grooving. oxDNA1.5 adds sequence-dependent interaction strengths to oxDNA1.0 \cite{Sulc2012Sequence-dependentModel}, and oxDNA2.0 \cite{Snodin2015IntroducingDNA} also includes a more accurate structural model, alongside an explicit term in the potential for screened electrostatic interactions between negatively charged sites on the nucleic acid backbone. In addition to these three versions of the DNA model, an RNA parameterisation ``oxRNA" has also been introduced \cite{Sulc2014ARNA}. 

In all three parameterisations, oxDNA represents each nucleotide as a rigid body with several interaction sites, namely the backbone, base repulsion, stacking and hydrogen-bonding sites, as shown in Fig. \ref{fig:interaction_sites}. In oxDNA1.0 and oxDNA1.5, these sites are co-linear; the more realisic geometry of oxDNA2.0 offsets the backbone to allow for major and minor grooving. 

Interactions between nucleotides depend on the orientation of the nucleotides as a whole, rather than just the position of the interaction sites. In particular, there is a vector that is perpendicular to the notional plane of the base, and a vector that indicates the direction of the hydrogen bonding interface. These vectors are used to modulate the orientational dependence of the interactions, which allows the model to represent the coplanar base stacking, the linearity of hydrogen bonding and the edge-to-edge character of the Watson–Crick base pairing.  Furthermore, this representation  allows the encoding of more detailed structural features of DNA, for example, the right-handed character of the double helix and the anti-parallel nature of the strands in the helix. 

The potential energy of the system is calculated as:
\begin{multline}
      V_{0}=\sum_{\langle ij \rangle}(V_{b.b.}+V_{stack}+V_{exc}^{\prime}) \\ +  \sum_{i,j \not\in \langle ij \rangle}(V_{HB}+V_{cr.st.}+V_{exc}+V_{coax}),
      \label{eq:potential}
\end{multline}
with an additional screened electrostatic repulsion term for oxDNA 2.0. In Eq.~\ref{eq:potential}, the first sum is taken over all pairs of nucleotides that are nearest neighbors on the same strand and the second sum comprises all remaining pairs. The terms represent backbone connectivity ($V_{b.b.}$), excluded volume ($V_{exc}$ and $V_{exc}^{\prime}$), hydrogen bonding between complementary bases ($V_{HB}$), stacking between adjacent bases on a strand ($V_{stack}$), cross-stacking ($V_{cr.st.}$) across the duplex axis and coaxial stacking ($V_{coax}$) across a nicked backbone. The excluded volume and backbone interactions are a function of the distance between repulsion sites. The backbone potential is a spring potential mimicking the covalent bonds along the strand. All other interactions depend on the relative orientations of the nucleotides and the distance between the hydrogen-bonding and stacking interaction sites.
  
  A crucial feature of the oxDNA model is that the double helical structure is driven by the interplay between the hydrogen-bonding, stacking and backbone  connectivity bonds. The stacking interaction tends to encourage the nucleotides to form co-planar stacks; the fact that this stacking distance is shorter than the backbone bond length results in a tendency to form helical stacked structures. In the single-stranded state, these stacks can easily break, allowing the single strands to be flexible. The geometry of base pairing with a complementary strand locks the nucleotides into a much more stable double helical structure.

  The model was deliberately constructed with all interactions pairwise (i.e., only involving two nucleotides, which are taken as rigid bodies). This pairwise character allows us to make effective use of cluster-move Monte Carlo (MC) algorithms, which provide efficient equilibrium sampling (see Section \ref{sec:parallel}).
  
  It is convenient to use reduced units to describe lengths, energies and times in the system. A summary of the conversion of these ``oxDNA units" to SI is provided in Appendix \ref{app:units}.                                                                                   
\begin{figure*}
       \includegraphics[width=\textwidth]{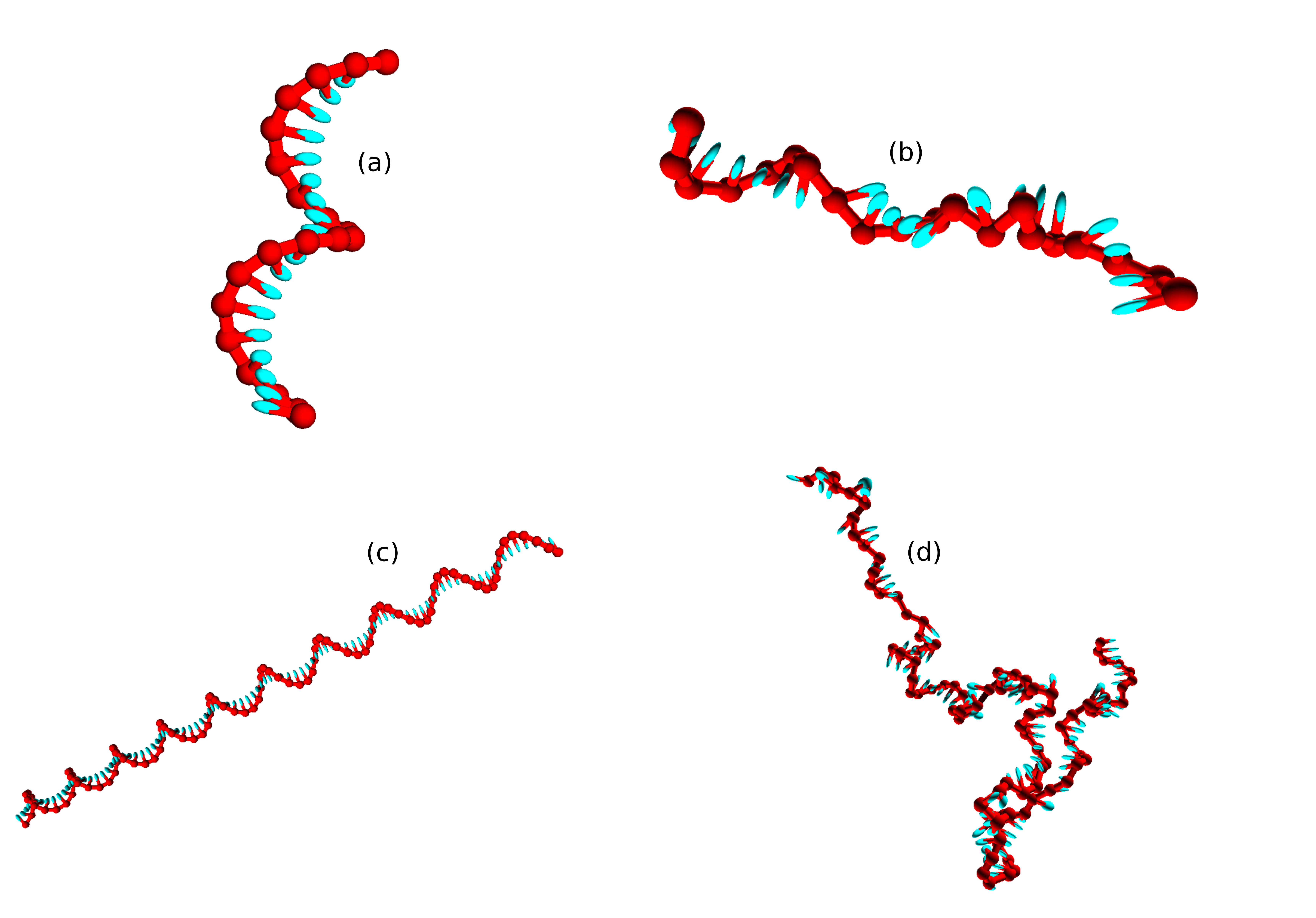}
        \caption{\small Snapshots of poly(dT) molecules used in the equilibration time tests. Non-representative initial states of poly(dT) molecules (left), and representative configurations obtained post-equilibration (right). (a) and (b): poly(dT) with 20 nucleotides; (c) and (d) poly(dT) with 100 nucleotides.} 
        \label{fig:polyA_ini_final}
    \end{figure*}

\section{Simulating the model: MD vs VMMC}
\label{sec:simulation}

\begin{figure*}[htbp]
    \includegraphics[width=\textwidth]{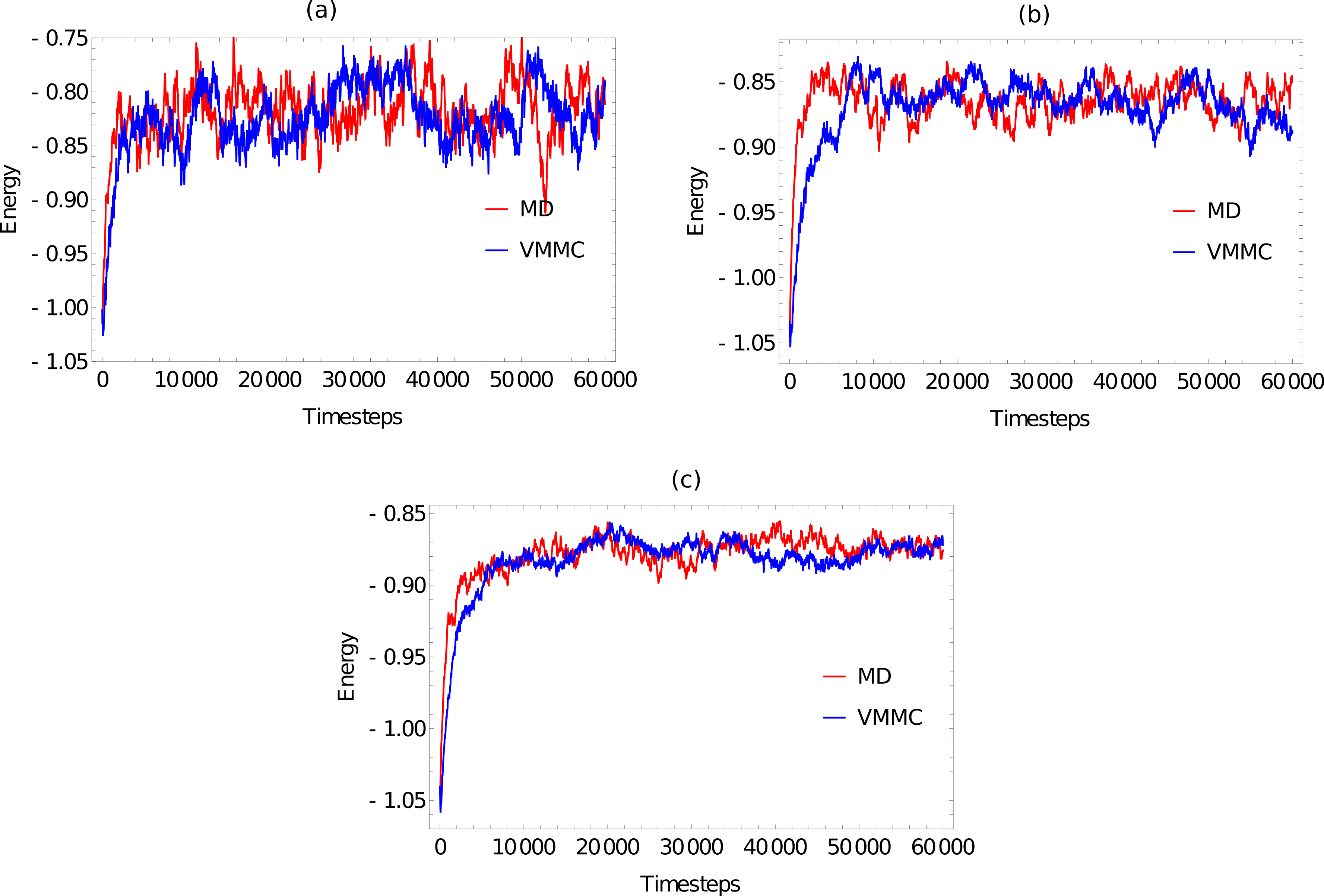}
     \caption{ \small Equilibration plots for the ploy(dT) molecules with (a) 20 bases, (b) 100 bases, (c) 1000 bases, obtained as averages over 20 independent simulations. For both MD and VMMC, the potential $V_0$ is plotted as a function of simulation progress, in units of reduced time (3.03\,ps) for MD and attempted steps per particle for VMMC.}
        \label{fig:MD_vs_VMMC}
\end{figure*}

The oxDNA model is far too complicated to approach analytically. Publicly released code to simulate oxDNA is available as a standalone package \cite{oxDNAwebsite}, or as a module \cite{oxDNALAMMPS} for the popular LAMMPS simulation software. There are two broad types of simulation technique that can be applied to probe the model: molecular dynamics (MD) and Monte Carlo (MC). 

Molecular dynamics  \cite{Frenkel2002UnderstandingApplications} algorithms evolve their constituent molecules according to Newton’s laws of motion, and so are a natural choice for simulating particle systems. For coarse-grained models such as oxDNA, in which the solvent is implicit, it is necessary to include a thermostat to both set the temperature and ensure diffusive rather than ballistic dynamics. The default MD algorithm for the standalone version of oxDNA is an Andersen-like algorithm \cite{Russo2009Reversible}, in which particle velocities and angular velocities are resampled from a Boltzmann distribution with a frequency that sets the effective diffusion coefficient. In the LAMMPS implementation, the model utilises a Langevin thermostat for rigid bodies \cite{Davidchack2014}, which applies small friction- and noise-based updates to the momentum and angular momentum at each step. The relative size of these contributions sets the temperature. 

A challenge of MD simulations is that when strong, short-ranged interactions are present -- as in oxDNA -- they place a limit on the maximum integration time step that can be used while preserving numerical stability. Interestingly both the Andersen-like and Langevin thermostats act to stabilise the simulations, allowing larger time steps to be used than if the equations of motion were integrated without noise or drag to generate energy-conserving, ballistic motion.

Both MD algorithms generate dynamical trajectories that can be used to probe system kinetics (more on this in Section~\ref{sec:dynamics}). However, it is also common to use MD to take equilibrium averages over the configurations of a particular system. In the limit of small time steps, both Andersen-like and Langevin algorithms will converge on a steady state in which they sample configurations ${\bf x}$ from the Boltzmann distribution $p_{\rm eq}({\bf x}) \propto \exp(-\beta V_0({\bf x})/k_B T)$, where $V_0({\bf x})$ is the potential energy of the model. How small the step size needs to be depends on a number of details, such as the strength of coupling to the thermostat. For parameters that have become an unofficial default for oxDNA, we illustrate the accuracy of the algorithm as a function of step size in Appendix~\ref{app:accuracy}. 

Monte Carlo (MC) \cite{Metropolis1949TheMethod} simulations are an alternative approach for sampling from the same Boltzmann distribution, but evading the drawbacks caused by the presence of a timestep altogether. In the standard MC approach \cite{Frenkel2002UnderstandingApplications}, configurational moves ${\bf x} \rightarrow {\bf y}$ are proposed randomly, with a symmetric probability distribution that satisfies $p_{\rm gen}({\bf x} \rightarrow {\bf y}| {\bf x}) = p_{\rm gen}({\bf y} \rightarrow {\bf x}| {\bf y})$. If these proposed moves are accepted with a probability $p_{\rm acc}({\bf x} \rightarrow {\bf y}) = {\rm max} \left(\exp(-(V_0({\bf y})-V_0({\bf x}))/k_BT \right)$, then the Boltzmann distribution $p_{\rm eq}({\bf x}) \propto \exp(-\beta V_0({\bf x}))$ is the stationary distribution of the simulation and a long simulation will sample from that distribution, assuming ergodicity. 

In principle, the moves ${\bf x} \rightarrow {\bf y}$ can be arbitrarily large without leading to errors, since it is not necessary to integrate the derivative of the potential, only calculate its values at the endpoints. However, standard MC techniques incorporate sequential updates of individual particles as the moves ${\bf x} \rightarrow {\bf y}$. For a model of a strongly-attractive system such as oxDNA, these moves must be extremely small or the acceptance factor will always be small. The result is painfully slow equilibration, particularly if large scale movements of strands is required to observe it. 

Virtual-move Monte Carlo (VMMC) \cite{Whitelam2009TheSelf-assembly,Whitelam2007AvoidingParticles} is an alternative that circumvents the drawbacks of MC algorithms. VMMC first proposes a single particle move, then generates a co-moving cluster of particles based on which interactions are best preserved by moving the particles in unison. The cluster building process is based on assessing the change in pairwise interactions, and so VMMC is especially suited to oxDNA, which has exclusively pairwise interactions. We have implemented the variant from the appendix of Ref.~\onlinecite{Whitelam2009TheSelf-assembly} in the standalone code. 

For those with limited experience of simulating oxDNA, it is not obvious whether VMMC or MD is the optimal approach to sampling a given system. We illustrate the relative efficiencies of the two algorithms when simulating ssDNA of length 20, 100 and 1000 bases, in terms of the computational time required to reach states representative of equilibrium from the same unrepresentative starting condition. 

We simulate the poly(dT) molecules with (a) 20 bases, (b) 100 bases and (c) 1000 bases, using oxDNA1.0  \cite{Ouldridge2011StructuralModel}.
Simulations are performed at $T = 27^{\rm o}$C, in a periodic box of 20, 100 and 1000 simulation units for 20, 100 and 1000 bases, respectively. For the MD simulations, we simulate for 60,000 simulation units of time (a nominal 182\,ns), with a time step of $d t=0.003$ (see Appendix \ref{app:accuracy}); each simulation therefore has $2 \times 10^7$ steps in total. For VMMC, we attempted 60,000 VMMC steps per particle. The proposed moves are: rotation about a random axis, through an angle up to 0.22 radians; and translation through a distance of up to 0.22 units. These choices produce a nice balance of cluster sizes, ranging from individual nucleotides to entire strands. 

Strands are initialized in a fully stacked, helical conformation as illustrated in Fig.~\ref{fig:polyA_ini_final} (a) and (c), and relax to more-representative, partially-stacked conformations (Fig.~\ref{fig:polyA_ini_final} (b) and (d)) as the simulation is run. The relaxation of the strands is associated with an increase in the potential $V_0$, and so we illustrate equilibration by plotting that potential averaged over 20 independent simulations as a function of simulation progress in Fig.~\ref{fig:MD_vs_VMMC}. The average value of the potential in equilibrium, $ \langle V_0 \rangle_{\rm eq}$, can be approximated by the average over the data collected in the second half of the simulations. We then estimate the equilibration time scale as the time required for $\langle V_0(t) \rangle-\langle V_0 \rangle_{\rm eq}$ to reach $1/e$ of its initial value for the first time. 

    \begin{table}[!]
\centering
\begin{tabular}{ |>{\centering\arraybackslash} p{1.2cm}| >{\centering\arraybackslash} p{1.2cm}| >{\centering\arraybackslash} p{1.9cm}| >{\centering\arraybackslash} p{1.9cm}| >{\centering\arraybackslash} p{1.9cm}| } 
 \hline
 \centering
 Strand length & Total runtime (MD) in seconds & Total runtime (VMMC)\; in seconds & Equilibration time as a fraction of total runtime (MD) & Equilibration time as a fraction of total runtime (VMMC) \\
 \hline
 \hline
 20 & 409.67 & 14.804 &  0.0118 & 0.0244 \\ 
 100 & 1804.18 & 147.57 & 0.0128 & 0.0287\\ 
 1000 & 32493.0 & 4187.03 & 0.0134 & 0.0299\\ 
  \hline
\end{tabular}
\caption{ Computational time for equilibration of poly(dT) molecules of various lengths. Simulations were performed using a single core Intel(R) Core(TM) i5-4300U CPU @ 1.90GHz.}
\label{table:polydt_equil_times}
\end{table}

The ``simulation progress" axes in Fig. \ref{fig:MD_vs_VMMC} are not directly comparable for MD and VMMC; one measures simulation time, the other attempted VMMC steps per particle. The most relevant quantity is the actual computational time required to equilibrate the system on a given architecture; in simple contexts, this time is also indicative of the speed with which the algorithm samples the equilibrium ensemble. Table \ref{table:polydt_equil_times} shows the total runtime of the simulations and the equilibration time as a fraction of that runtime. In computational time, the VMMC algorithm is able to equilibrate the poly(dT) molecules more quickly (compared to MD algorithm) in all the three cases. VMMC is around 15 times as fast for the 20-nucleotide strand, dropping to around 4 times as fast for the 1000-nucleotide strand. The large moves available to VMMC, and the lack of a requirement to differentiate potentials, provide this benefit. Note, however, that the ratio of the equilibration times for MD to VMMC algorithms decreases as the system size increases.

The relative efficiency of VMMC and MD approaches will depend to some degree on the choice of damping parameters and seed moves; we have not carefully optimised our choices for either technique, but have used values that generally work well. The relative efficiency will also depend on the particular system: VMMC lends itself to systems in which large movements are important. Nonetheless, the general rule of thumb that VMMC is more efficient for smaller systems - particularly those with significantly fewer than 1000 nucleotides - is a helpful one. It is also particularly easy to enhance VMMC using umbrella sampling, as explained in Section \ref{sec:umbrella}. 

For sufficiently large systems, such as DNA origami, MD should equilibrate faster, and therefore provide improved sampling. Another major advantage of the MD approach is much more facile parallelisation when simulating large systems. The standalone code allows for parallel simulation on GPUs (graphical processing units), and the LAMMPS module for parallel simulations across multiple CPUs (central processing units) using MPI. These approaches are demonstrated in Section \ref{sec:parallel}.

\section{Mechanical properties of DNA}

The mechanical properties of DNA are central to its role across nanotechnological, biophysical and biological contexts. DNA's flexibility and response to applied stress determine the conformation and accessibility of the genome inside cells \cite{lewis1996crystal,nikolov1996crystal,richmond2003structure,widom2001role}. Moreover, not only is the stiffness of dsDNA important in maintaining the conformation of DNA nanostructures, but the relative flexibility of ssDNA crucially allows for joints and flexible hinges. These properties are widely-studied in bulk and single-molecule experiments {\it in vitro} \cite{crothers19921,fujimoto2006torsional,bryant2003structural,wang1997stretching,Smith1996OverstretchingMolecules,gross2011quantifying,gore2006dna,lionnet2006wringing,Mills1999,Seol2007StretchingBase-stacking,Rivetti1998,Dessinges2002,Seol2004ElasticRibonucleotide,le2014probing,podtelezhnikov2000multimerization,du2008kinking,demurtas2009bending,kim2015dynamic,fields2013euler,allemand1996elasticity,forth2008abrupt,mosconi2009measurement,brutzer2010energetics,salerno2012single,tempestini2013magnetic}.

It is therefore essential that a coarse-grained model provides a reasonable representation of these properties. In this section, we both discuss the mechanical properties of oxDNA, and show how to construct simulations that can probe these properties. 

\subsection{Stiffness of duplex and single-stranded DNA}
The most common metric used to quantify the stiffness of DNA is the persistence length, defined in the textbook of Cantor and Schimmel as \cite{Cantor1980BiophysicalMacromolecules}
\begin{equation}\label{eq:persistence_length}
L_{\rm ps}=\frac{ \langle \mathbf{L} \cdot \mathbf{l}_{0}\rangle }{\langle l_{0} \rangle}.
\end{equation}
Here, $\mathbf{L}$ is the end-to-end vector of the polymer and $\mathbf{l}_{0}$ represents the vector between the first two monomer units. dsDNA is most commonly thought of as a semi-flexible polymer or wormlike chain \cite{kratky1949rontgenuntersuchung}. In this picture, the discrete series of inter-base pair vectors are approximated as a continuous, differentiable polymer axis with a quadratic free energy of curvature. For an infinitely long, semi-flexible polymer, correlations in the alignment of the polymer axis decay exponentially with separation, with a decay rate given determined by $L_{\rm ps}$. When translated back to the language of inter-base-pair vectors, we obtain
\begin{equation}\label{eq:persistence_exponential}
    \frac{\langle \mathbf{l}_n \cdot \mathbf{l}_{0}\rangle}{\langle l_0 \rangle^2} = \exp(-n \langle l_{0}\rangle/L_{ps}),
\end{equation}
where ${\bf l}_n$ is the vector between base pair $n-1$ and base pair $n$.

It is relatively straightforward to both assess whether the wormlike chain model is a good model for oxDNA, and to extract $L_{\rm ps}$. We simply simulate a duplex system for long enough to sample a representative set of configurations, calculate the correlation between inter-base-pair vectors as a function of separation, and fit the results to the exponential decay of Eq.~\ref{eq:persistence_exponential}. 

\begin{figure}[h]
            \includegraphics[width=8cm]{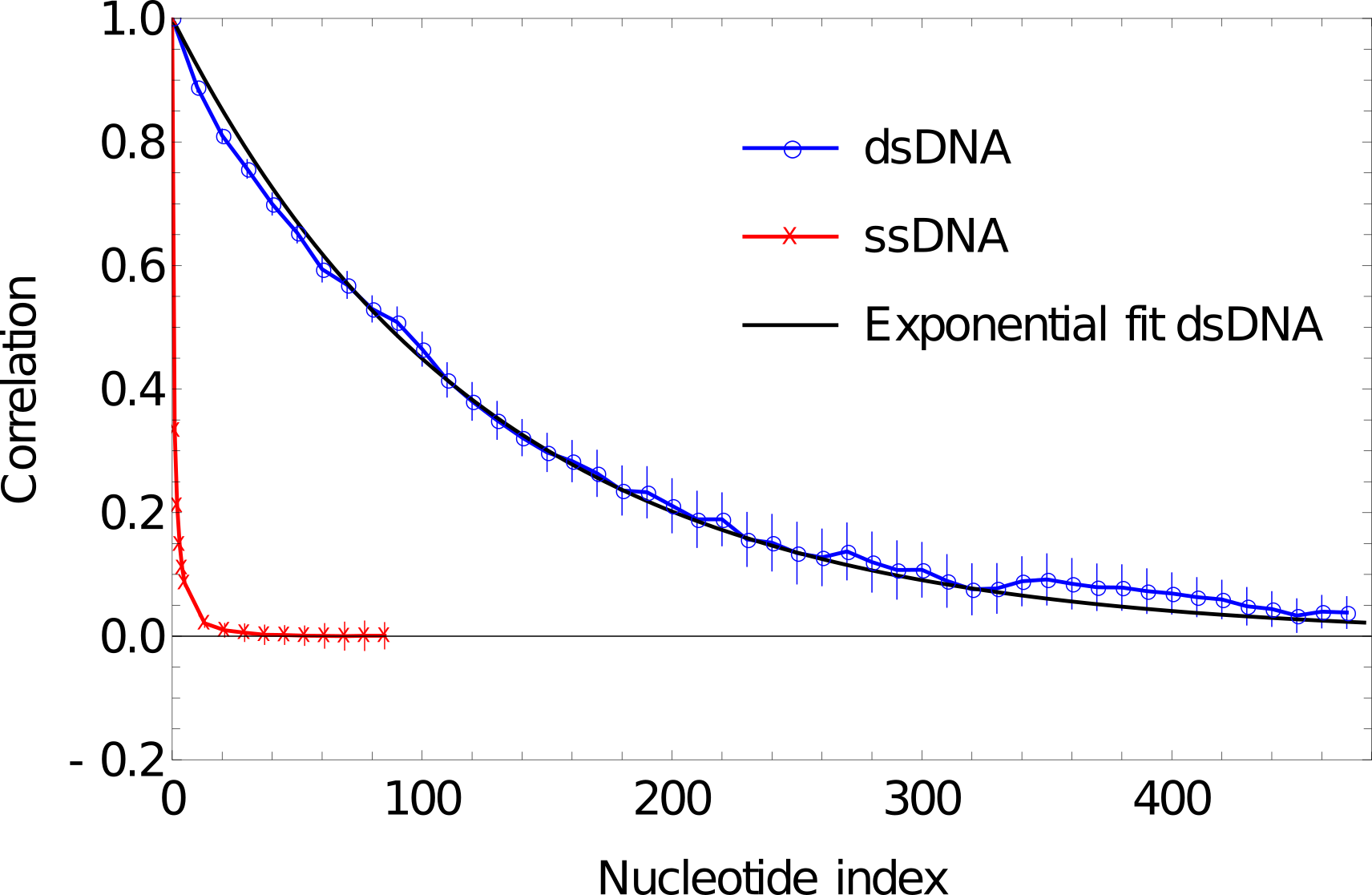}
            \caption{ Plot of correlation of inter-segment vectors vs distance (number of base pairs along the DNA) for 500 dsDNA (blue curve) and 100ssDNA (red curve). dsDNA fitted with an exponential decay with a decay constant of 0.0076134.}
            \label{fig:correlation}
\end{figure}

\begin{figure}[h]
            \includegraphics[width=8cm]{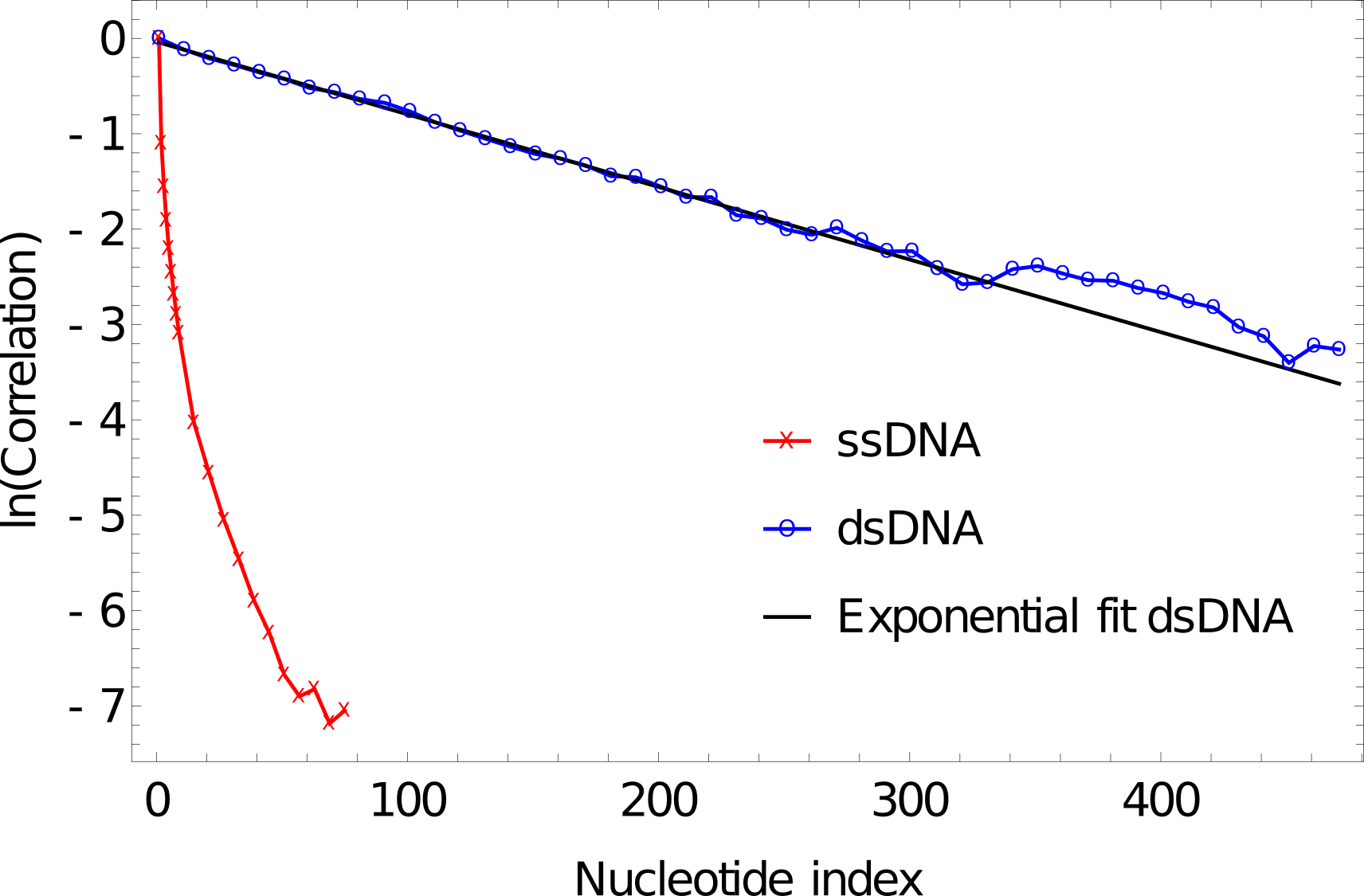}
            \caption{Log plot of correlation of inter-segment vectors vs distance (number of base pairs along the DNA) for 500 dsDNA (blue curve) and 100ssDNA (red curve). dsDNA fitted with an exponential decay with a decay constant of 0.0076134.}
            \label{fig:lncorrelation}
\end{figure}

In Fig.~\ref{fig:correlation}, we plot the results of such a procedure. To obtain these data, we simulate a DNA duplex of length 500 base pairs at 27$^\circ$C using oxDNA1.5. We perform 20 VMMC simulations with $2 \times 10^7$ attempted steps per particle for VMMC. The first $10^5$ moves are treated as an initialization period and no data is collected. Additionally, the base pairs at the ends of the duplex are more flexible than those well within the bulk; to obtain properties representative of bulk DNA, we therefore do not include the five base pairs at either end in our analysis.  Correlations are calculated from the default configurational outputs of the model, using the code provided in \cite{supportfiles}. In order to obtain a good sample, it is helpful to output these configurations with a high frequency; we use a small value for the parameter to output energy configurations after a single VMMC move per particle.

As is evident from Fig.~\ref{fig:lncorrelation}, the correlation of the duplex axis indeed follows an exponential fall-off, to within sampling error. Fitting eq.~\ref{eq:persistence_exponential} to $\ln \left(\frac{\langle \mathbf{l}_n. \mathbf{l}_{0}\rangle}{\langle l_0 \rangle^2}  \right)$
gives $L_{\rm ps} \approx 131 \langle l_0 \rangle = 131\times 0.4118 \approx 54.758 = 46.6nm $, consistent with experimental estimates of 40-50\,nm (120-150 base pairs) at high [Na+] concentrations \cite{harrison2015coarse,harrison2019identifyingmodel}.

Indeed, more generally, the mechanical properties of double-stranded oxDNA are well-described by a semiflexible polymer model, and its torsional and extensional moduli have been analysed elsewhere \cite{Ouldridge2011StructuralModel,Matek2015}. Significant deviations from this behaviour - such as sharp kinks facilitated by broken base pairs - are generally only observed when large stresses are applied to the molecule \cite{Romano2013Coarse-grainedOverstretching,Matek2015}, in agreement with experiment.   

ssDNA behaves very differently in oxDNA. In Fig. Fig.~\ref{fig:correlation}, we plot the correlations of backbone-site-to-backbone-site vector for a 100-base poly(dT) ssDNA, obtained from running simulation in oxDNA1.5  at 27$^\circ$C. We perform 20 VMMC simulations with $2\times 10^{7}$ attempted steps per particle. The first $10^{5}$ moves are treated as an initialization period and no data is collected.  For these simulations, we have set the stacking strength between the nucleotides to zero (the consequences of non-zero stacking strength will be addressed in Section \ref{sec:f-e}).  From Fig.~\ref{fig:correlation} and Fig.~\ref{fig:lncorrelation}, it is apparent that the correlation drops very rapidly, meaning that unstacked ssDNA is very flexible in oxDNA, as it should be; adjacent backbone-to-backbone vectors can bend through a large angle. But importantly, it is worth noting that the drop in correlation between vectors with separation along the polymer cannot be well described by an exponential as in eq.~\ref{eq:persistence_exponential}. The convexity of $\ln \left(\frac{\langle \mathbf{l}_n. \mathbf{l}_{0}\rangle}{\langle l_0 \rangle^2}\right)$ is indicative of more distant backbone-to-backbone vectors being aligned more strongly than would be expected from the alignment of two adjacent backbone-to-backbone vectors.

The reason for this behaviour is that it is the excluded volume of nucleotides that gives unstacked ssDNA its ``stiffness'' in oxDNA. The excluded volume of nucleotides discourages ssDNA from folding back on itself, but importantly it leads to very different polymer properties than assumed in common polymer models such as the freely-jointed chain and the wormlike chain. For these classic polymer models, the statistical properties are entirely determined by interactions between parts of the polymer that are adjacent along the backbone, whereas the curvature of  $\ln \left(\frac{\langle \mathbf{l}_n. \mathbf{l}_{0}\rangle}{\langle l_0 \rangle^2}\right)$  in Fig.~\ref{fig:lncorrelation} is indicative of interactions between more distant points along the polymer contour playing a role. 

As a result, using a wormlike chain with a given $L_{ps}$ (or a freely jointed chain with a given Kuhn length) to understand ssDNA in oxDNA is misleading. The overall tendency of the polymer to swell to fill a large volume - due to its excluded volume - would suggest a far greater degree of local stiffness than actually present.  This effect is retained even when stacking between adjacent nucleotides is included. 

Importantly, these complexities also apply to physical ssDNA, as well as the oxDNA model. Single deoxyribonucleotides have linear dimensions on the order of 1\,nm, and experimental attempts to measure the mechanical properties of oxDNA (usually reported as persistence lengths) are of a similar order of magnitude \cite{Mills1999,Rivetti1998,Murphy2004,Smith1996OverstretchingMolecules}. Describing ssDNA in this way is not self-consistent; any polymer with this cross-section and flexibility would be strongly affected by excluded volume, so these models cannot be accurate. The result has been that experiments on large scale properties of relaxed ssDNA \cite{Rivetti1998,Murphy2004}, which are sensitive to excluded volume effects, tend to produce larger estimates for quantities like $L_{ps}$ than experiments on shorter sections of ssDNA, or sDNA under high tension \cite{Smith1996OverstretchingMolecules,Rivetti1998}. 

Low salt concentrations, which lead to weaker screening of electrostatic interactions between non-adjacent nucleotides make the above effect stronger \cite{Smith1996OverstretchingMolecules,Dessinges2002}. Base-pairing interactions in non-homopolymeric ssDNA have a confounding effect; the formation of secondary structure tends to condense the strand, making it appear more flexible when its statistics are modelled with a wormlike chain or a freely-jointed chain \cite{Smith1996OverstretchingMolecules}. Overall, as for oxDNA, simple descriptions of the mechanical properties of physical ssDNA should be treated with caution.

\subsection{\label{sec:f-e}Response of ssDNA to tension}
\begin{figure*}
    \includegraphics[width=\textwidth]{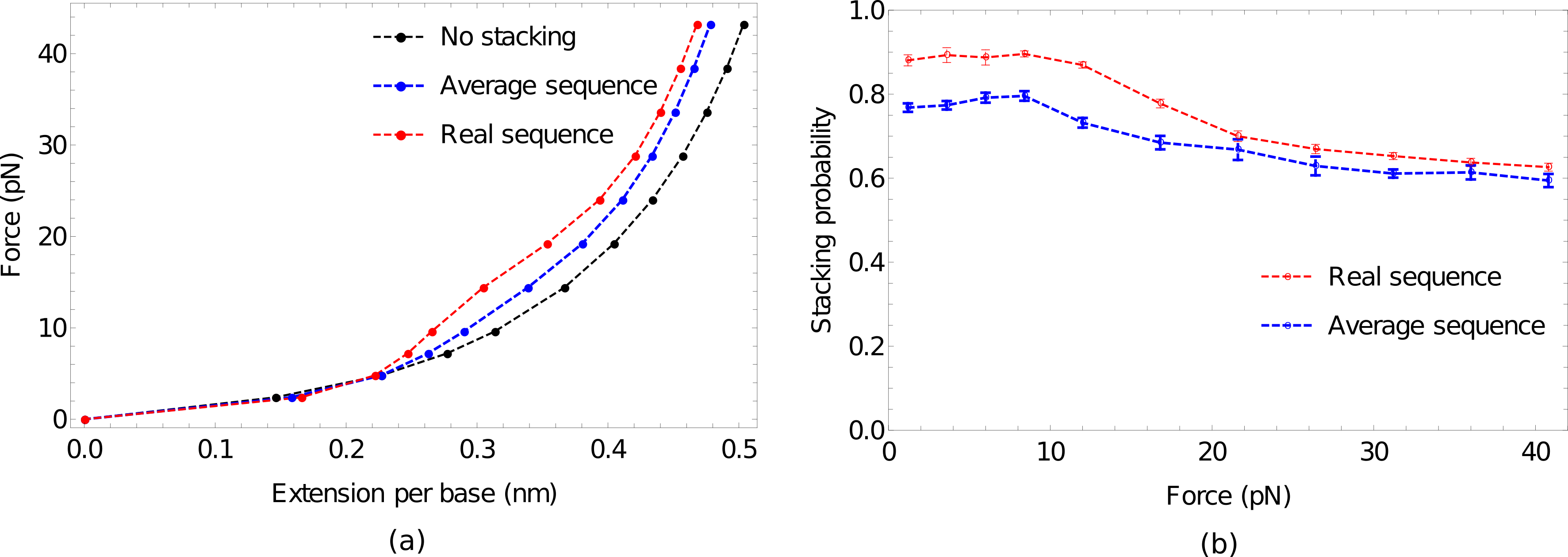}
  
        \caption{\small The response of ssDNA to tension, and the role of stacking therein. (a) Force-extension plots for 100-nucleotide poly(dA) using three models: no stacking (black); average stacking strength (oxDNA1.0, blue) and sequence-dependent stacking (oxDNA1.5, red). Stronger stacking leads to an increased force at larger extensions, and extremely strong stacking results in a plateau-like feature as stacking is disrupted. (b) Stacking probability for average stacking strength (oxDNA1.0, blue) and sequence-dependent stacking (oxDNA1.5, red) as a function of applied force. Adjacent nucleotides are defined as stacked if the stacking energy between the pair is more than -0.1 units. } 
    \label{fig:force_extension}
\end{figure*}

A common mechanism for probing the mechanical properties of DNA is to apply force, whether torsional \cite{bryant2003structural,allemand1996elasticity,forth2008abrupt,mosconi2009measurement,brutzer2010energetics,salerno2012single,tempestini2013magnetic}, extensional \cite{Huguet2010,Smith1996OverstretchingMolecules,wang1997stretching,gross2011quantifying,gore2006dna,Seol2007StretchingBase-stacking,Dessinges2002,Seol2004ElasticRibonucleotide,allemand1996elasticity} or shearing \cite{hatch2008demonstration,wang2013defining,van2009unraveling,forth2008abrupt,mosconi2009measurement,brutzer2010energetics,salerno2012single,tempestini2013magnetic}. 

Experiments have focused on both the elastic properties associated with small deformations \cite{bryant2003structural,wang1997stretching,Smith1996OverstretchingMolecules,allemand1996elasticity}, and large scale structural transitions \cite{gross2011quantifying,gore2006dna,hatch2008demonstration,wang2013defining,van2009unraveling,bryant2003structural,Smith1996OverstretchingMolecules,Seol2007StretchingBase-stacking,Dessinges2002,Seol2004ElasticRibonucleotide,allemand1996elasticity,forth2008abrupt,mosconi2009measurement,brutzer2010energetics,salerno2012single,tempestini2013magnetic}.

Applying external tension is relatively straightforward in molecular simulation; there are more subtleties associated with applying boundary conditions for external torsion \cite{Matek2012,Matek2015}, but it is also possible. In the case of oxDNA, small external stresses have been used to help parameterise and characterise the model \cite{Ouldridge2011StructuralModel,Matek2015,Nomidis2019Twist-bendBeyond,skoruppa2017dna};
 larger stresses have been applied to provide insight into experiments on structural transitions \cite{Romano2013Coarse-grainedOverstretching,Matek2012,Matek2015,mosayebi2015force,engel2018force,desai2020coarse,Wang2015Twist-inducedDensity,wang2014modeling}.

Systems with internally-induced stress, where the drive to form base pairs in one part of an assembly applies stress to another part, have also been studied \cite{harrison2019identifyingmodel,harrison2015coarse,fosado2021twist,park2021sequence,caraglio2019overtwisting,tee2018well,wang2017influence,Sutthibutpong2016Long-rangeSimulation,wang2016sequence,engel2020measuring}.

As an example, in this section we demonstrate the force-extension properties of ssDNA as represented by oxDNA. Optical tweezer experiments with ssDNA have a long history \cite{Smith1996OverstretchingMolecules}. These original experiments with naturally-occurring DNA exhibited formation and stabilization of secondary structure in high salt conditions and low-moderate force, although this was not explicitly modelled at the time. The presence of this secondary structure makes simulation of DNA heteropolymers hard; it is challenging to equilibrate a long strand with many competing base-pairing configurations (we have had some success using methods based on parallel tempering \cite{Romano2013Coarse-grainedOverstretching}). Instead, therefore, we simulate 100-nucleotide-long homopolymeric poly(dA) using oxDNA1.0 and oxDNA1.5. 

The helicity in oxDNA is driven by stacking interactions between adjacent nucleotides. As is evident from Fig.~\ref{fig:polyA_ini_final}, this stacking has a residual effect on the structure of ssDNA strands, which are partially stacked in equilibrium. We will use force-extension simulations to probe the consequences of single-stranded stacking in oxDNA.

For these simulations,  which are similar to original results in \cite{Sulc2012Sequence-dependentModel},  we use both a version of the parameters with no stacking interacting, a sequence-averaged stacking interaction (oxDNA1.0), and a sequence-specific stacking interaction (oxDNA1.5) for which poly(dA) has the strongest interaction of all sequences. All simulations are performed at $T = 27^{\rm o}$C, in a periodic box of length 100 simulation units with each simulation running for $4\times8$ steps with $d t=0.005$ which equals to a total run time of $2\times 10^6$ units. 4 sets of simulations are performed for 12 difference values of force.

Fig.~\ref{fig:force_extension}(a) shows that extensive stacking has only a moderate effect on the force-extension properties of the ssDNA at low force. In the sequence-dependent model, poly(dA) is close to 90\% stacked at 27$^\circ$ C - see Fig.~\ref{fig:force_extension}(b). However, the increased stiffness due to the tendency to form stacked single helices (akin to the initial state in Fig.~\ref{fig:MD_vs_VMMC}) is counteracting by the shorter end-to-end distance of the backbone when it is forced to wind around the helix. 

At larger forces, however, we clearly see a signal of stronger stacking. Larger force is required to extend the strands with stronger stacking, and a plateau-like feature is evident in the system with the strongest stacking. A similar plateau was  observed by Seol \emph{et al.} for RNA stretching (poly(A) and poly(C)) \cite{Seol2004ElasticRibonucleotide,Seol2007StretchingBase-stacking} but was absent for poly(U). Those authors hypothesised that the plateau arises as the shorter end-to-end distance in helical stacked confirmations becomes prohibitive; additional force is then required to disrupt the stacking interaction to allow further extension, causing an increase in the gradient of the force-extension curve. After the bases have unstacked, the gradient becomes less steep again. 

Broadly speaking, this explanation is borne out by oxDNA. Notably, however, Seol  \emph {et al.}  \cite{Seol2004ElasticRibonucleotide,Seol2007StretchingBase-stacking} concluded that a relatively low stacking probability should give a pronounced plateau. By contrast, in Fig.~\ref{fig:force_extension}(b) -- obtained by probing configuration output files to assess the degree of stacking \cite{supportfiles} -- we see that strands with an initial stacking probability of 78\% show only a hint of the plateau. This discrepancy arises because, in the minimal model of Seol  \emph {et al.} \cite{Seol2004ElasticRibonucleotide,Seol2007StretchingBase-stacking}, even a single pair of stacked nucleotides has a much shorter end-to-end distance along the ssDNA backbone than an unstacked pair. The explicit representation of 3D structure in oxDNA, however, captures the fact that the shortening of the end-to-end distance along the DNA backbone is only significant when several bases in a row are stacked into a helix, so that the backbone really has to wrap back round upon itself. As a result, extension can occur whilst disrupting only a fraction of the stacking interactions, and ssDNA in oxDNA remains significantly stacked even at high force (Fig.~\ref{fig:force_extension}(b)). 

Whilst oxDNA's representation of the polynucleotide backbone is simplistic, these geometrical arguments also apply to physical DNA - suggesting that even weak plateau-like behaviour in ssDNA force-extension curves is evidence of strong stacking, and the absence of a plateau is not proof of an absence of stacking. More generally, this system is indicative of the value that oxDNA can provide. The system involves an interplay between basic structure, mechanics and thermodynamics of ssDNA. When applied, oxDNA reveals subtleties that are not directly apparent from a more minimal model. Indeed, it is quite common to construct very simple models to interpret biophysical experiments on the mechanical properties of DNA \cite{Vafabakhsh2012,qu2011complete,fields2013euler,salerno2012single,tempestini2013magnetic,meng2014coexistence,hatch2008demonstration,wang2013defining}; simulations with oxDNA often reveal physically reasonable relaxation mechanisms that aren't factored into these simpler models \cite{Matek2015,harrison2019identifyingmodel,harrison2015coarse,mosayebi2015force,skoruppa2017dna}. At this stage, it is also worth highlighting a general virtue of coarse-grained models that is apparent in these simulations. It is very simple just to switch off interactions -- such as the stacking here -- to isolate the effect those interactions have on the system. Doing so can be incredibly helpful in interpreting the physical cause of experimental signals.

\section{Thermodynamic simulations with oxDNA}
\label{sec:thermo}
\subsection{Duplex formation thermodynamics}
As well as representing the structure and mechanical properties of ssDNA and dsDNA, oxDNA is also designed to capture the thermodynamics of the hybridization transition from ssDNA to dsDNA. Needless to say, accurately capturing the thermodynamics of this transition is essential for any model hoping to describe biological and nanotechnological processes involving the forming and disruption of base pairs.

To assess the thermodynamics of a simple duplex, it is typical to simulate an isolated pair of strands in a periodic cell that is large enough to prohibit self-interactions (unit cell size of $\gtrapprox 2n$ oxDNA length units, where $n$ is the duplex length, is generally sufficient). Given a sufficiently long VMMC or MD simulation, the fraction of time spent in the bound state can be estimated and used to infer quantities such as melting temperatures, as outlined below. 

However, particularly for longer strands, simulating this process can be prohibitively slow.  For two short strands in solution, the vast majority of configurations have well-separated strands and no base-pairing interactions. Enthalpically favourable base-pairing provides a compensatory advantage to configurations with many well-formed base pairs (fully-formed duplexes). To obtain a good estimate of the fraction of strands bound in equilibrium, it is necessary to pass between these two sub-ensembles (completely unbound and fully bound) many times; as a rule of thumb, we have found that around 10 interconversions will start to provide meaningful statistics. 

Unfortunately, interconversion requires the system to transition through states with only one or two base pairs that benefit neither from the large ensemble of configurations accessible to dissociated strands, nor the favourable interactions of fully-bound strands. These configurations with $Q=1,2$ base pairs are rare in the equilibrium ensemble, and have a relatively high free energy
\begin{equation}\label{eq:free_energy}
    F(Q)=-kT\ln\left(p_{\rm eq}(Q) \right)  +C, 
\end{equation}
where $C$ is a $Q$-independent constant, and $p_{\rm eq}(Q)$ is the probability of observing $Q$ base pairs in equilibrium.  The high free energy of these intermediate states makes dissociation and association rare event processes that are challenging to sample directly.

\subsubsection{Umbrella Sampling}
\label{sec:umbrella}
To overcome this difficulty, simulations can be augmented with umbrella sampling \cite{Torrie1977}. For a system with coordinates $x$ (in our case, nucleotide positions and orientations), umbrella sampling involves identifying a collective order parameter $\lambda(x)$ for the transition of interest, and then applying a bias $W(\lambda(x))$ to force the system to occupy otherwise undesirable values of $\lambda(x)$ that lie along the transition path more frequently. Unbiased statistical averages can be extracted from these biased samples using
\begin{equation}\label{eq:biased_sampling}
    \langle A(x) \rangle_{\rm eq} = \left\langle \frac{A(x)}{W(\lambda(x))} \right\rangle_{\rm biased},
\end{equation}
where $A(x)$ is a quantity of interest. Essentially, the contribution of each configuration sampled to the average is reduced by a factor of the bias applied. 

A common approach with umbrella sampling is to perform a series of separate simulations with very strong biases tightly centred on distinct values of $\lambda(x)$. Simulations centred on adjacent values of $\lambda(x)$ can then be knitted together using procedures such as the Weighted Histogram Analysis Method (WHAM), allowing the calculate of the free energy difference between the start and end point \cite{Kumar1992}. 

\begin{figure}[h]
            \includegraphics[width=8cm]{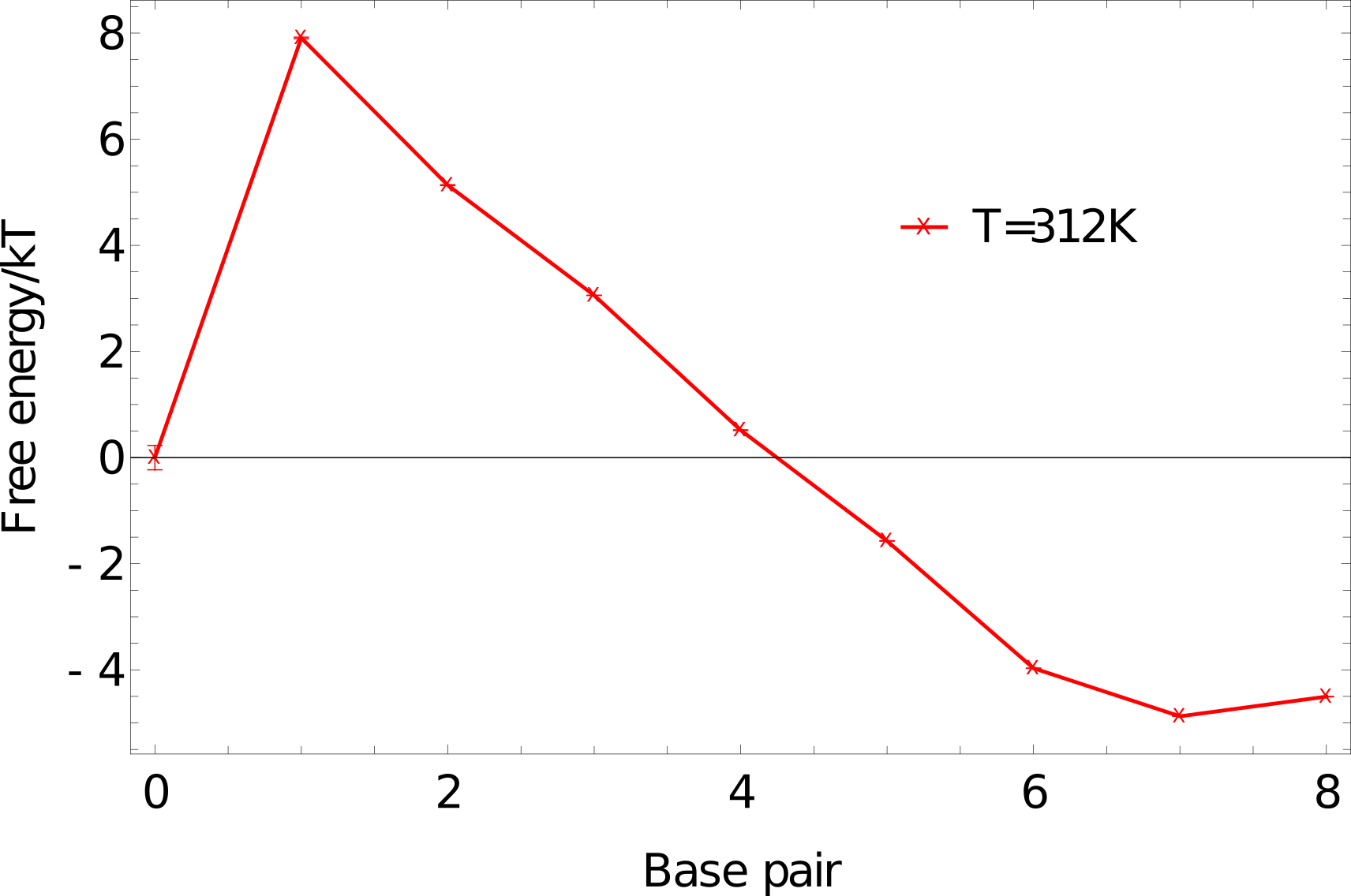}
            \caption{Free-energy profile of an 8-base-pair duplex (3'-ACTGACGT-5' and 3'-ACGTCAGT-5') at 312K in a simulation volume of side length 15 units. }
            \label{fig:8bp_energy}
\end{figure}

Generally, however, we have found that this sophisticated approach is not necessary for oxDNA, and a particularly straightforward umbrella sampling method is built into the standalone oxDNA code. When using VMMC, it is possible to specify discrete order parameters $\lambda (x)$ based on the number of base pairs between user-defined groups of nucleotides. For duplex formation, it is fairly straightforward to iteratively identify a biasing potential that facilitates both the sampling of all states and the rapid transition between fully bound and completely detached configurations. 

This biasing potential doesn't need to be fine tuned so that all values of $\lambda(x)$ are equally probable in the biased sample -- just good enough to facilitate multiple transitions backwards and forwards. Typical examples for 5-base and 8-base duplexes are given in \cite{supportfiles}. For more complex systems, more sophisticated $\lambda(x)$ and the use of multiple sampling windows are sometimes necessary. Even in these cases, however, the principles are similar to those outlined here. 

We perform umbrella sampling simulations on an 8-nucleotide duplex at 312K in a simulation volume of side length 15 units, using the oxDNA1.5 version of the model. 5 independent simulations are performed for $7.7\times10^8$ VMMC steps per particle. The quantity $p_{\rm eq}(Q)$ obtained from simulations is used to calculate a free energy $F(Q)$ according to Eq.~\ref{eq:free_energy} and plotted in Fig.~\ref{fig:8bp_energy}. The shape of this graph is typical for duplex formation, showing the expected large jump in free energy from 0 to 1 base pairs. From 1 to 6 base pairs there is a steady drop in the free energy as configurations are stabilised by additional base-pairing interactions that are favoured once the strands are in close proximity. The final base pairs are less favourable, as base pairs at the end of a duplex are prone to fraying \cite{SantaLucia2004TheMotifs,Ouldridge2011StructuralModel}.

The shape of $F(Q)$ gives a good guide to constructing first estimates of umbrella biases $W(\lambda(x))$ for duplex formation in general. Ignoring the dissociated state, $W(\lambda(x))$ should increase roughly exponentially with the number of base pairs broken, since it must counteract $\exp(-F(Q)/kT)$. The slope of $F(Q)$, and hence the required rate of exponential growth in  $W(\lambda(x))$, is determined by the temperature; as a crude rule of thumb, a bias of a factor of 10-15 is required per base pair broken at 300K; this required bias falls to a factor of 3-4 by 330K. 

The initial jump in free energy from 0 to 1 in Fig.~\ref{fig:8bp_energy} is largely determined by the simulation volume; for simulation cells similar in size to this one, a factor of 3000-10000 is a reasonable first guess for the required weight of the 1-base-pair state relative to the 0-base-pair state. 

In addition to biasing by the number of base pairs formed, it is sometimes helpful to also use a distance-based contribution to the order parameter. Built in to the standalone oxDNA code is the ability to define additional dimensions of $\lambda(x)$ that depend on the minimum separation between sets of nucleotides, rather than the number of base pairs. We have found that a simple division of the 0-base-pair state into configurations in which the strands are close (less than 4 units apart) and far apart (4 or more units apart) can reduce the amount of time spent sampling the independent diffusion of strands around the simulation volume. For simulation volumes similar to this one, the close state should be weighted by around 5-10 relative to the distant state, which dominates the unbound ensemble. 

\subsubsection{Melting temperature curves}

\begin{figure}[h]
            \includegraphics[width=8cm]{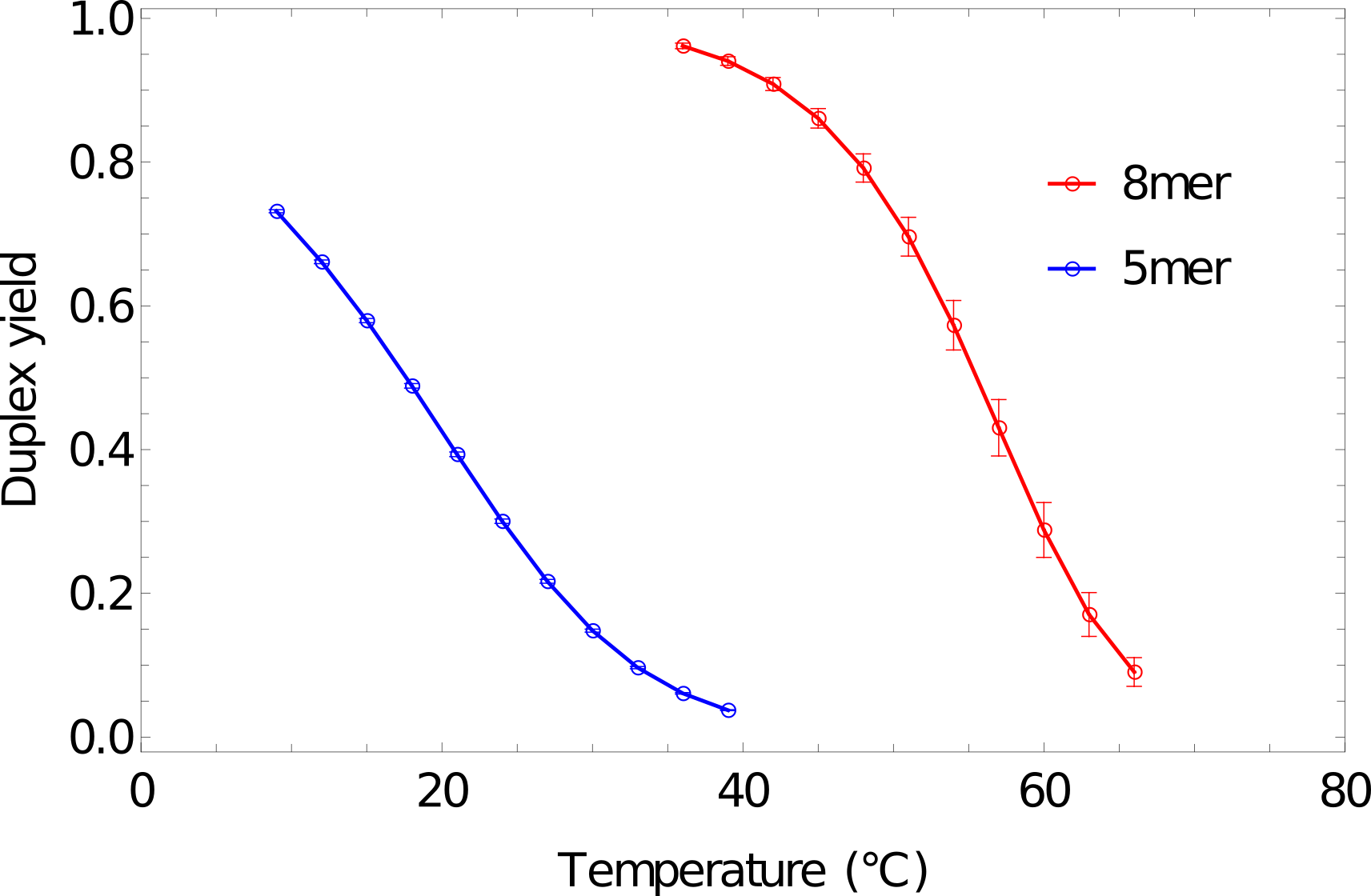}
            \caption{Melting transition of oligonucleotides. Fractional yield of 5mer (3'-AGTCT-5'/3'-AGACT-5') and 8mer (3'-ACTGACGT-5'/3'-ACGTCAGT-5') duplxes i bulk ($f_\infty(T)$)) as predicted by oxDNA1.5 at a total concentration of $7.96 \times 10^{-4}$\,M for each strand.}
            \label{fig:melting_curve}
\end{figure}

Although free-energy profiles for a single pair of strands are informative, they aren't directly comparable to the majority of experiments. Indeed, the thermodynamics of the oxDNA model was parameterised to reproduce the nearest neighbour model \cite{SantaLucia1998AThermodynamics}, which in turn was fitted to -- and predicts -- experimental melting curves in bulk conditions. The Santalucia parameterisation of the nearest neighbour model \cite{SantaLucia2004TheMotifs} assumes that DNA duplex formation is essentially a two-state transition between a well-formed duplex and separated single strands. The free-energy profiles produced by oxDNA, such as Fig.~\ref{fig:8bp_energy} are consistent with this picture; the ensemble is dominated by configuration with either zero base pairs, or a large number. In this limit, the melting behaviour can be well-characterised by the fraction of strands that are expected to have base pairing with another strand at a temperature $T$ in a bulk system, $f_{\infty}(T)$.

To calculate $f_{\infty}(T)$ in this two-state description, it is first necessary to obtain data at a range of temperatures. In principle, these data can be obtained through separate simulations. However, we have found that a technique called single histogram reweighting \cite{Ferrenberg1988} is sufficient to infer $f_{\infty}(T)$ accurately over a large enough range of temperatures to describe the melting transition. The basic idea is to treat a simulation at a temperature $T$ as a biased sample of the ensemble at another temperature $T^\prime$; this bias can be corrected in the same way as the bias applied during umbrella sampling:  
\begin{equation}\label{eq:biased_sampling2}
    \langle A(x) \rangle_{T\prime} = \langle A(x)\exp{\left(V_0(x,T)/kT - V_0(x,T^\prime)/kT^\prime\right)} \rangle_{T}.
\end{equation}
Here $V_0(x,T)$ is the value of the potential in the original simulation at temperature $T$ ($V_0(x,T^\prime)$ is slightly different due to a $T$-dependent term in the potential \cite{Ouldridge2011StructuralModel}). Extrapolation to nearby temperatures using single histogram reweighting is built into the oxDNA standalone code. It is important to note that if umbrella sampling, and particularly temperature reweighting, are applied, then it is especially important to simulate for a good equilibration time before results are collected. Normally,  any initial unrepresentative states will be swiftly overwhelmed within an average taken over the whole course of the simulation. The unbiasing factors in Eq.~\ref{eq:biased_sampling} and Eq.~\ref{eq:biased_sampling2}, however, can cause unrepresentative initial states to be assigned enormous weights in the ensemble average that are effectively insurmountable, rendering the simulation results meaningless.

Given well-sampled data of the formation of a single duplex in a simulation volume, it is tempting to assume that the fractional yield of states with more than one base pair in a single duplex simulation, $f_1(T)$, is equal to the bulk yield of duplexes $f_\infty(T)$ in a system with the same total concentration of strands. Unfortunately this is not the case; simulations of only a single target duplex neglect concentration fluctuations within unit cells that have large effects on the yield of products \cite{Ouldridge2010ExtractingSimulations,Ouldridge_bulk_2012}. Quantitative comparison to experimental data is therefore impossible unless extrapolations to bulk conditions can be performed. Assuming ideal behaviour of solutes, Extrapolation is possible. For dimerisation between non-self-complementary strands \cite{Ouldridge2010ExtractingSimulations}
\begin{equation}
    f_\infty(T) = \left(1 + \frac{1}{2 \Phi(T)} \right) - \sqrt{\left(1 + \frac{1}{2 \Phi(T)} \right)^2-1}.
\end{equation}
where $\Phi(T) = f_1(T) /(1-f_1(T))$. A similar result holds for self-complementary duplexes \cite{Ouldridge2010ExtractingSimulations}, and algorithms exist to extrapolate to bulk for more complex assemblies \cite{Ouldridge_bulk_2012}.

Melting curves obtained for 5-base and 8-base duplexes, using umbrella sampling, temperature reweighting and extrapolation to bulk, are reported in Fig.~\ref{fig:melting_curve}. The melting temperatures $T_m$ for these duplexes -- defined, in the two state model, as the temperature at which $f_\infty(T)$ is 0.5 -- are close to the values predicted by the nearest neighbour model at the same conditions (17.8\degree C and 56.1\degree C) \cite{SantaLucia2004TheMotifs}. This agreement is, of course, due to the model being fitted to these data. However, it is worth noting that although duplexes are often described as having a single ``melting temperature", the  temperature at which $f_\infty(T)$ is 0.5 depends on the concentration of the individual strands, $[C]$ with \cite{Ouldridge2011StructuralModel} 
\begin{equation}
    \frac{{\rm d} T_m}{{\rm d}[C]} \sim \frac{\Delta T}{[C]}.
\end{equation}
Here, $\Delta T$ is the width of the transition over which $f_\infty(T)$ goes from largely bound to largely unbound. To match nearest neighbour predictions for melting temperatures over a range of concentrations, therefore, it is necessary that transition widths are also comparable; achieving a good match was a major part of oxDNA's parameterisation. 

\begin{figure*}
    \centering
       \includegraphics[width=\textwidth]{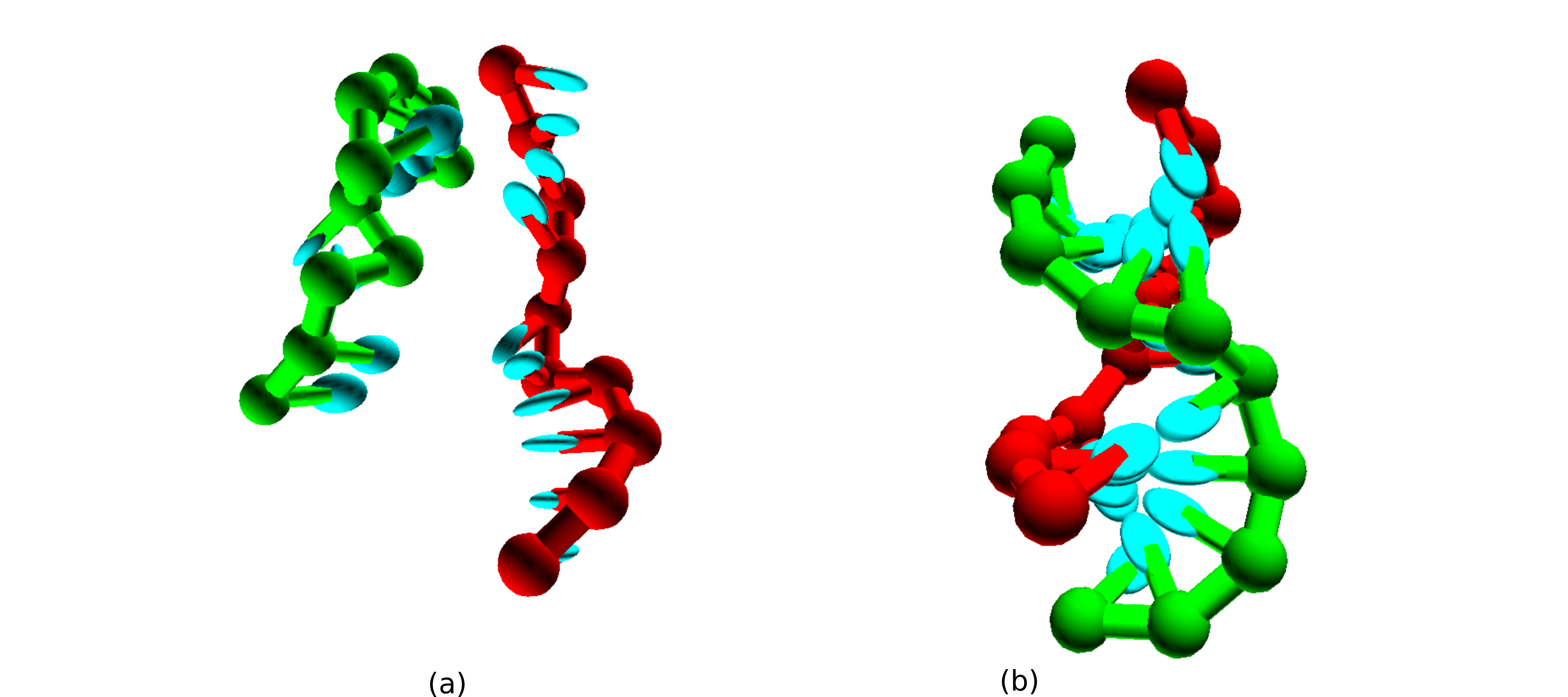}
         \caption{\small Pseudoknot unbound (a) and bound (b) states, for seqeunces 3'-AGCTTCCATG-5'/3'-AAGCTCATGG-5'.  } 
    \label{fig:pseudoknot_pictures}
\end{figure*}

\begin{figure}[h]
            \includegraphics[width=8cm]{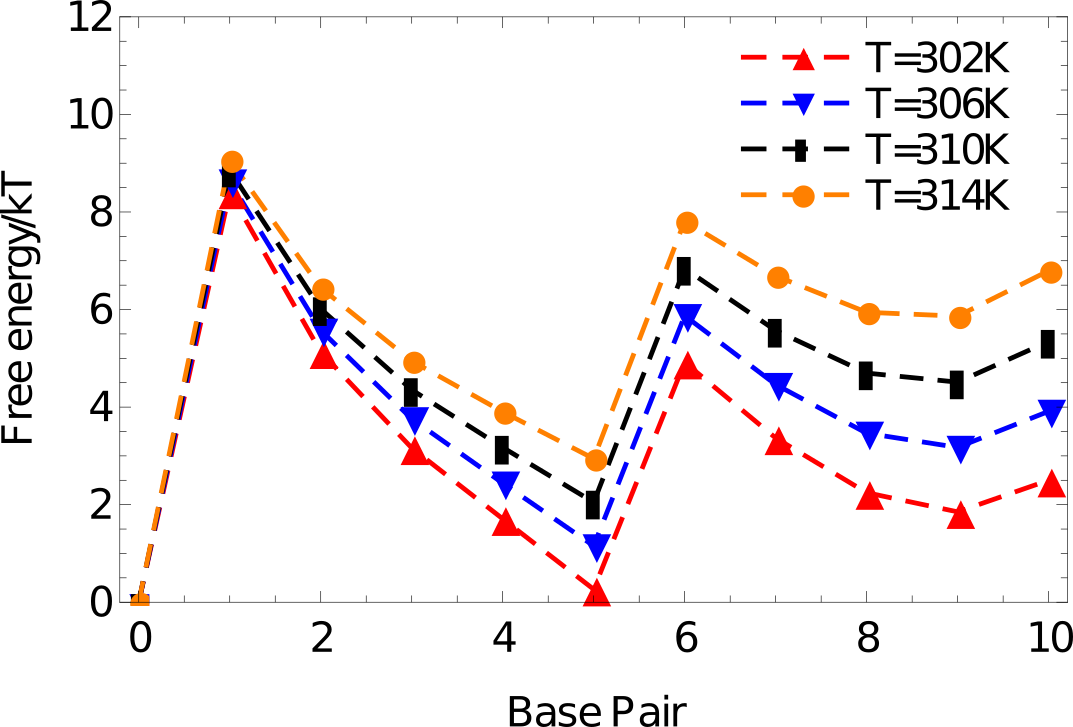}
            \caption{Free energy vs number of base pairs formed for the complex 3'-AGCTTCCATG-5'/3'-AAGCTCATGG-5' in a simulation volume of side length 20 oxDNA units.}
            \label{fig:pseudoknot_energy}
\end{figure}

\subsection{Thermodynamics of more complex structures}

Although accurately simulating basic duplex formation was necessary for the parameterisation of oxDNA, little new information is to be gained from performing these simulations again. The model is trained to reproduce the thermodynamics of the nearest neighbour model, so simulating the thermodynamics of duplex formation is an expensive way to get at an approximation to said nearest neighbour model. 

Where oxDNA can add value is if duplex formation occurs as part of some more complex system - possibly one in which internally or externally-applied stresses, or topological constraints, are relevant \cite{Ouldridge2013OptimizingWalker,Sulc2012Sequence-dependentModel,Romano2012TheComplexes,harrison2019identifyingmodel,mosayebi2015force,tee2018well,kovcar2016design}. As an example, we simulate the formation of a small pseudoknotted structure (Fig.~\ref{fig:pseudoknot_pictures}) leveraging the intuition and techniques discussed in Section \ref{sec:thermo}. Here, the two sequences 3'-AGCTTCCATG-5' and 3'-AAGCTCATGG-5' cannot form a single continuous duplex, but can form two 5-bp duplexe section if both strands bend back on themselves. The stability of this structure cannot be inferred from the nearest neighbour model, but it can easily be simulated with oxDNA. Applying umbrella sampling, we simulate the system at a temperature of  308K in a periodic cell of side-length 20 using oxDNA1.5. 

The resulting free-energy profile, Fig. \ref{fig:pseudoknot_energy}, shows that, at the temperatures of interest, forming two arms is less favourable than forming only one. The advantage obtained by bringing the strand into close proximity via the binding of the first duplex is not enough to overcome the internal stress generated by the structure. This internal stress is evidenced by the much shallower slope of the free energy profile for forming base pairs 6-10 than 1-5.

    \begin{figure*}[htbp]
        \centering
        
            \includegraphics[width=\textwidth]{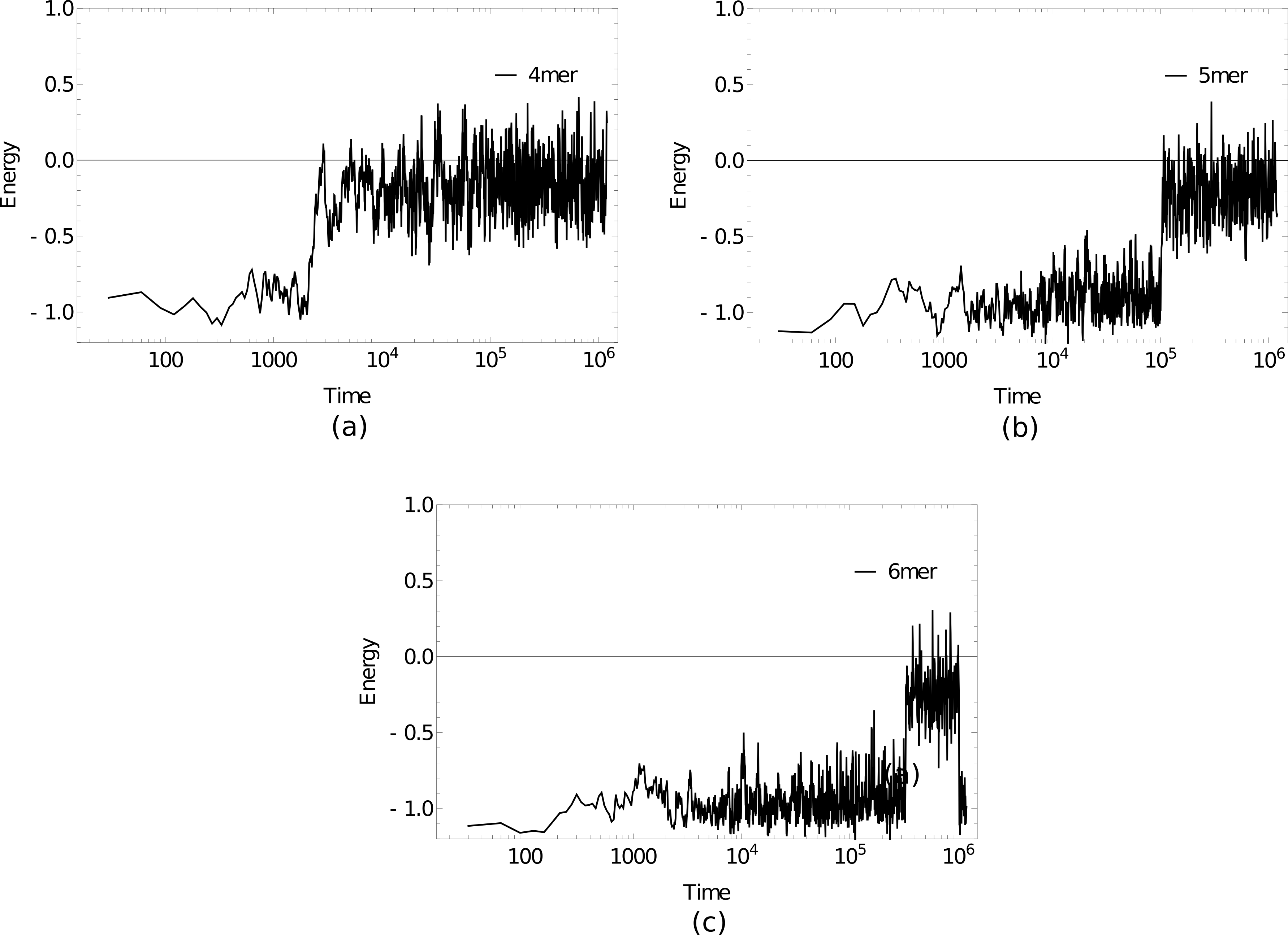}
            \caption{ \small Energy per nucleotide vs simulation time step for (a) 4mer, (b) 5mer, (c) 6mer duplexes at 320K. Sudden transitions from low to high energy are indicative of rare-event melting.}
        \label{fig:MD_melting}
\end{figure*}

\section{Dynamical simulations}
\label{sec:dynamics}

The simulations described hitherto probe static quantities obtained in the equilibrium ensemble. However, the dynamics of DNA-based systems can be equally important. In particular, the time required for reactions to happen is crucial when constructing complex self-assembling systems or functional circuits, particularly those that are intended to remain out of equilibrium, or exhibit an extremely slow relaxation to equilibrium \cite{dunn2015guiding,cabello2021handhold,srinivas2017enzyme,fern2018modular}.

Unlike the thermodynamic and structural properties, oxDNA has not been carefully parameterised to the dynamics of physical DNA. Coarse-graining is generally known to speed up timescales by smoothing free-energy landscapes \cite{murtola2009multiscale}. Moreover, the explicitly dynamical algorithms (particularly the Andersen-like thermostat) give a fairly crude approximation to the dynamics expected from small molecules in solution. Neither the Anderson-like nor the Langevin thermostat incorporates cooperative hydrodynamics (an updated version of the Langevin thermostat developed to describe hydrodynamic effects \cite{davidchack2017geometric} has not yet been implemented in LAMMPS), and both are typically run with large effective diffusion coefficients to enhance sampling (see Appendix \ref{app:units}).

Nonetheless, the dynamics of oxDNA is fundamentally constrained by the combination of its free-energy landscape and its embedding of that free-energy landscape in an explicit geometrical description. For comparison, it is surprisingly difficult to generate meaningful dynamics based on just the free-energy landscape predicted by the nearest neighbour model without an explicit geometrical representation \cite{Srinivas2013,schaeffer2015stochastic} [add commented-out citation].

As a result, dynamical simulations of oxDNA can provide useful insight into dynamical properties of physical DNA; the model has been particularly successful in describing toehold-mediated strand displacement \cite{Srinivas2013,Machinek2014ProgrammableDisplacement,haley2020design,irmisch2020modeling}, one of the fundamental reactions of DNA nanotechnology. Importantly, the focus should always be on comparing the relative dynamics of two similar systems -- for example, the dependence of strand displacement rates on toehold lengths. Unlike the thermodynamic and mechanical properties of oxDNA, absolute values of dynamical properties are largely irrelevant.

As an example, we simulate the dissociation kinetics of duplexes of length 4 (3'-ATAT-5'/3'-ATAT-5'), 5 (3'-ATATA-5'/3'-ATATA-5') and 6 (3'-ATATAT-5'/3'-ATATAT-5') at 320K using the Anderson-like thermostat applied to oxDNA1.5. Example trajectories, showing the energy of the system per nucleotide, illustrate the two-state nature of the system discussed in Section \ref{sec:thermo}. The strands spend a substantial amount of time in states with an energy of approximately -1.0 in oxDNA units (duplex configurations), before suddenly transitioning to states with an energy around -0.2 (single-stranded states). As hinted at by these examples, longer strands take exponentially longer to dissociate (the simulation steps taken to reach an energy of -0.2 per nucleotide, averaged over 10 simulations for each length, are: $9117\pm2883$, $64032\pm20249$, $279744\pm88463$). This exponential suppression of the dissociation rate with strand length is consistent with dissociation being a rare event that requires the crossing of a free energy barrier whose height grows linearly with duplex length, as suggested by the free-energy profile in Fig.~\ref{fig:8bp_energy}.

In this case, all systems studied showed the required behaviour on relatively short time scales. Frequently, it is necessary to simulate much slower processes. We have found that the forward flux sampling (FFS) technique \cite{Allen2009} is an effective tool for simulating dynamical processes with a longer timescale. However,  FFS is trickier to implement than umbrella sampling, and is not yet built in to the released code in an optimal way.

\section{Simulation of large structures}
\label{sec:parallel}
Another significant application area for oxDNA has been the simulation of large structures to assess their conformation, stability and flexibility \cite{berengut2020self,yao2020meta,Poppleton2020DesignSimulation,schreck2016characterizing,berengut2019design,hoffecker2019solution,tortora2020chiral,choi2018design,brady2019flexibility,Snodin2019Coarse-grainedOrigami,benson2018effects,sharma2017characterizing,shi2017conformational,fernandez2016small,chhabra2020computing,coronel2018dynamics}. 

In this context, oxNDA represents an alternative to the CanDo model and simulation package \cite{Castro-cando-2011}. The added complexity of oxDNA has a computational cost, but means that it is better able to handle irregular systems. For such simulations, use of oxDNA2.0 is strongly recommended given its better representation of structure, particularly in the context of DNA origami. A more detailed primer on setting up these simulations can be fund in Ref.~\cite{DoyeOrigami2020}; here we focus only on technical aspects of the simulations.

As briefly mentioned in Section \ref{sec:simulation}, MD algorithms can facilitate the simulation of really large systems by allowing parallelisation across GPU threads or multiple CPUs. The oxDNA standalone code is GPU-enabled via the CUDA C API and supports runs on single CPUs and single GPUs, whereas the LAMMPS version of oxDNA uses the Message Passing Interface (MPI) and is optimised for parallel runs on multi-core CPUs and distributed memory architectures.

To provide benchmarks and examplar codes, we have performed large-scale simulations with both implementations on two different compute architectures, namely a NVIDIA V100 PCIe GPU with 5,120 CUDA cores at Arizona State University's High Performance Computing Facility, and the ARCHIE-WeSt HPC facility at the University of Strathclyde consisting of 64 Intel Xeon Gold 6138 (Skylake) processors @2.0GHz with 40 cores per node and 2,560 cores in total. The GPU and single-core CPU runs were performed with the oxDNA standalone code SVN version 6989. The GPU runs all used mixed precision \cite{Rovigatti2015JCC} and an edge-based approach \cite{Russo2011JCP}. The LAMMPS stable version from 3rd March 2020 was used for the multi-core CPU runs. All runs were performed with the oxDNA2.0 model featuring sequence-dependent stacking and hydrogen-bonding interactions. 

Two different benchmarks were studied to analyse the performance of both implementations. The first one consisted of a varying number of double-stranded octamer duplexes and investigated the performance at different system sizes, ranging from 8 octamers with 128 nucleotides in total to 262,144 octamers with 4,194,304 nucleotides in total. The concentration of octamers was kept constant at one octamer per 20$^3$ oxDNA length units, whereas the temperature and salt concentration were set to $T=293K$ and $[{\rm Na}^+]=500$\,mM, respectively. 

The second benchmark consisted of a DNA origami ``pointer" structure \cite{bai2012cryo} (15,238 nucleotides) and tested the performance at different salt concentrations between $[{\rm Na}^+] = 100$\,mM and 1\,M. The salt concentration is another performance-critical aspect in the simulation of nucleic acids that is often neglected. The reason is that the salt concentration affects the Debye screening length, which is proportional to the inverse square root of the salt concentration. The temperature of this second benchmark was fixed at $T=293K$. The initial configuration was converted from the cadnano format using the TacoxDNA server \cite{Suma2019TacoxDNA:Origami}, then relaxed using oxdna.org, implementing the protocol from \cite{Doye2020arXiv} followed by a simulation at the respective salt concentration. 

It is worth emphasising that origami structures such as the pointer are a setting in which the improved structural model of oxDNA2.0 is essential. Unless an accurate model is used, relatively small discrepancies can contribute strain that builds up across the structure, resulting in large scale distortion.

\begin{figure}[htbp]
        \centering
        \includegraphics[width=0.5\textwidth]{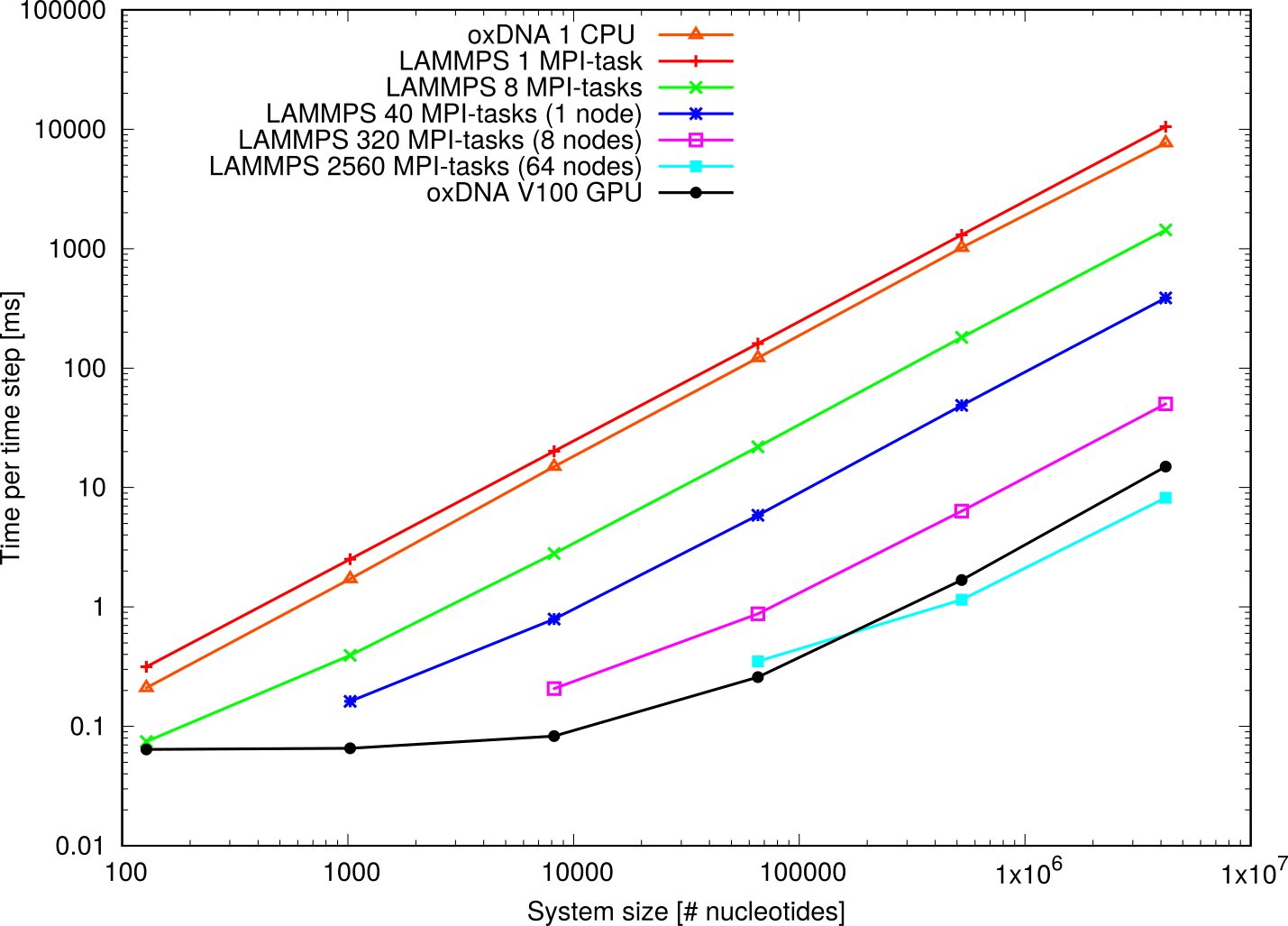}
        \caption[]
     { \small  Performance of the oligomer benchmark as time per time step for various system sizes: Shown are results of the oxDNA standalone code on a single CPU and NVIDIA V100 GPU and of the LAMMPS implementation of the oxDNA2 model at different CPU counts.}
        \label{fig:oligomer_scaling}
\end{figure}

Fig. \ref{fig:oligomer_scaling} shows the results of the oligomer benchmark, which are expressed as time per integration time step in milliseconds. On a single Intel Xeon Gold CPU the standalone code implements a single timestep slightly faster than the LAMMPS implementation. Note, however, that the actual efficiency will depend on the choice of coefficients of coupling to the thermostats \ref{app:accuracy}.

When deployed in parallel on more CPUs, the LAMMPS implementation offsets this disadvantage almost immediately. Its performance at the larger side of system sizes is more or less ideal as evidenced through the linear increase of time per integration step with system size. For smaller system sizes, and depending on how many CPUs were used, the performance levels off due to a build-up of MPI communication overheads. However, there is still a noticeable speed-up e.g. for 8,192 nucleotides on 320 MPI-tasks or 65,536 nucleotides on 2,560 MPI-tasks, which comes down to a very low 25 nucleotides per MPI-task. This unusually good performance of a parallel molecular dynamics code has been reported before \cite{Henrich2018Coarse-grainedApplications} and is owed to the rather complex oxDNA force field as the code spends a good deal of time carrying out the force calculation.

The GPU-implementation of the standalone code retains a significant advantage over the LAMMPS implementation for all but the largest benchmark sizes and runs on the full ARCHIE-WeSt system size (2,560 MPI-tasks) and its performance levels only off when the GPU becomes under-subscribed with threads at smaller system sizes. We can conclude that the LAMMPS implementation of oxDNA, besides its capability to run on a variety of CPU architectures, is very suitable for studying small and intermediate system sizes, whereas the GPU-implementation has clearly the edge at large-scale simulations.

\begin{figure}[htbp]
        \centering
        \includegraphics[width=0.5\textwidth]{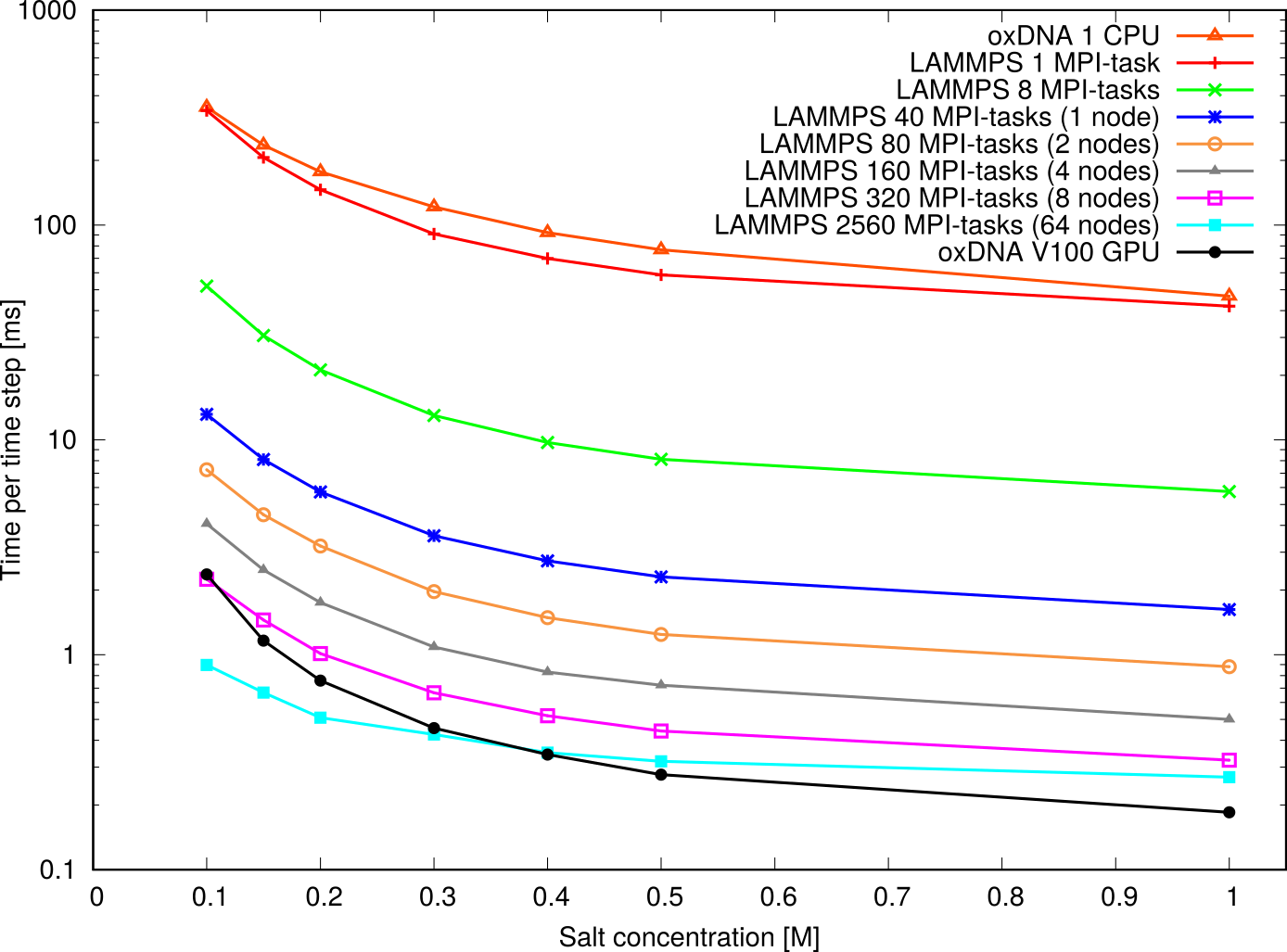}
        \caption[]
     { \small  Performance of the pointer benchmark as time per time step at various salt concentrations: Shown are results of the oxDNA standalone code on a single CPU and NVIDIA V100 GPU and of the LAMMPS implementation of the oxDNA2.0 model at different core counts.}
        \label{fig:pointer_scaling}
\end{figure}

Fig. \ref{fig:pointer_scaling} shows the performance with the pointer benchmark, again expressed as time per time step in milliseconds. This time the LAMMPS implementation is marginally faster than the oxDNA standalone code on a single CPU. Again, the GPU-runs of the standalone code features significantly shorter run times on all but the largest core counts and lowest salt concentrations. It appears the increase in run time between high and low salt concentration is slightly larger for the GPU-implementation of the standalone code. This could be due to a slightly better handling of neighbour lists in LAMMPS.

Most importantly, however,  an increase in runtime by a factor 8-9 can be seen at all core counts when moving from high to moderate salt concentrations. This slowdown is in line with the increase in Debye length by about a factor 3 and reflects the longer cutoff radii and neighbour lists of the pair interactions. This large performance difference should be taken into account when choosing simulation parameters: For instance it is nearly always more convenient to perform relaxation runs to create well-initialized configurations at high salt concentrations (\textit{e.g.} $[{\rm Na}^+] = 1$\,M). Indeed, unless the response of the system to decreased salt concentration is of specific interest, we would generally recommend using high monovalent salt concentration such as  $[{\rm Na}^+] = 1$\,M for the actual data collection.
     
\section{Conclusions}
We have reviewed the properties of, and simulation methods available for, the oxDNA model. In the process we have created a well-documented library of examplar simulations available from \cite{supportfiles}. Equally importantly, however, we have attempted to provide the necessary intuition both for successfully running oxDNA-based simulations, and also for identifying which systems would actually benefit from those simulations in the first place.

Having explored the model's strengths in some detail, it is worth noting a few natural directions for improvements. Although the model has well-parameterised sequence-dependent thermodynamics, and a good representation of average mechanical properties, it lacks sequence-dependent structure and mechanics. Incorporating this feature would be useful in and of itself, but would also be a useful first step towards building a model that could interface with other molecules such as proteins \cite{Procyk2020Coarse-grainedNanotechnology}.

\section*{Acknowledgements}
This work is part of a project that has received funding from the European Research Council (ERC) under the European Union’s Horizon 2020 research and innovation programme (Grant agreement No. 851910). T.E.O. is supported by a Royal Society University Fellowship.  O.H. acknowledges support from the EPSRC Early Career Research Software Engineer Fellowship Scheme (Grant No. EP/N019180/2).
This work used the ARCHIE-WeSt High-Performance Computer (www.archie-west.ac.uk) based at the University of Strathclyde.

\appendix
\section{Reduced units in oxDNA}
\label{app:units}
Input and output files use a specific set of reduced units (often called Lennard-Jones units) to represent the system. These units are referred to as ``simulation units'' in this review. Both the standalone implementation and the LAMMPS module use the same reduced units for lengths and energies.
\begin{itemize}
\item In the model, one unit of length $\sigma$ corresponds to 8.518 {\AA}. This value was chosen to give a rise per bp of approximately 3.4 {\AA} and equates roughly to the diameter $d$ of a nucleotide.
\item One unit of energy is equal to $\varepsilon$ = 4.142$\times 10^{-20}$ J (or equivalently,
$k_BT$ at $T$ = 300 K corresponds to 0.1$\varepsilon$). 
\item The cutoff for treating two bases as ``stacked'' or ``base-paired'' is conventionally taken to be when the relevant interaction term is more negative than -0.1 in simulation units (4.142$\times 10^{-21}$ J).
\item The units of energy and length imply a simulation unit of force equal to  4.863$\times 10^{-11}$\,N.
\end{itemize}

For the purpose of dynamical simulations, it is necessary to define a simulation unit of mass. For historic reasons, the standalone code and the LAMMPS code use slightly different definitions.

\vspace{3mm}

\noindent In the LAMMPS code,
\begin{itemize}
\item 1 simulation unit of mass $M$ is taken to be 100 AMU or $1.661\times 10^{-25}$ kg with a mass of $M=3.1575$ per nucleotide in simulation units. 
\item When combined with the simulation units for energy $\epsilon$ and length $\sigma$, the mass unit $M$ implies a simulation unit of time $\tau=\sigma\sqrt{M/\varepsilon}$ of 1.706 ps.
\end{itemize}

\noindent In the standalone code, 
\begin{itemize}
\item 1 simulation unit of mass $M$ is $5.243\times 10^{-25}$\,kg and nucleotides are assumed to have a mass $M=1$ in simulation units.
\item The simulation unit of time $\tau$ is therefore 3.031\,ps.
\end{itemize}

It is worth noting that both approaches treat the nucleotide as a sphere for the purpose of evaluating its equation of motion. The LAMMPS code uses a moment of inertia of 0.435179 in simulation units; the standalone code uses 1 in simulation units.

In dynamical simulations, it is also necessary to set the diffusion coefficient $D$ or the mobility $\mu$. For a sphere, both are related through the Einstein-Smoluchowski equation $D=\mu\,k_B\,T$, whereas the mobility is given through the Stokes-Einstein relation $\mu=1/(3\pi\,\eta\,d)$ with $\eta$ as solvent viscosity and $d$ the diameter. Typical values for aqueous solutions are $\eta=10^{-2}$ Poise $=10^{-3}$ kg m$^{-1}$s$^{-1}$ or 8.749 $M \sigma^{-1} \tau^{-1}$ in LAMMPS simulation units. 

It is instructive to determine the inertial and Brownian timescale from these values. The inertial timescale $\tau_{in}=1/\gamma$ is the timescale on which momentum relaxation occurs, and is the inverse of the friction constant $\gamma = 1/(m \mu)$ with $m$ as mass of the inert object. This leads to $\tau_{in}=3.829\times 10^{-2} \tau$ in LAMMPS simulation units. The Brownian timescale $\tau_B=d^2/D$ is the time it takes an object undergoing Brownian motion to diffuse over its own diameter. At $T=300$ K we obtain $\tau_B=8.246\times 10^2$ $\tau$, therefore $\tau_{in}\simeq 4.644\times 10^{-5}\tau_B\ll \tau_B$, which means both timescales are separated by more than five orders of magnitude.  

To improve the sampling efficiency and to maximise the actual physical time of a simulation, it is well justified to speed up the diffusion and opt for larger simulation values of $D$ and $\mu$ provided the sequence of timescales is not violated. This means $\tau_{in}=1/\gamma < d^2\,m\,\gamma/k_B T= d^2\,m / (k_B T \tau_{in}) = \tau_B$. In LAMMPS units this leads to $\tau_{in}<\sqrt{3.1575/k_B T}$ or $\gamma>\sqrt{k_B T/3.1575}$, so for instance at $T=300$ K $\tau_{in}<5.619$ in simulation units. Setting $\tau_{in}=2.5$ entails $\tau_B\simeq 5~\tau_{in}$ and results in inertial and Brownian timescales that are sufficiently separated.

\FloatBarrier
\section{Accuracy of thermostats}
\label{app:accuracy}
MD simulations only sample from the correct Boltzmann distribution in the limit of small integration timesteps. In this section, we illustrate the performance of the integrators in terms of reproducing the correct average energies. 
For the purposes of these simulations, we consider a small duplex with an overhanging single-stranded tail, simulated using oxDNA1.0 at 300 K. Sequences:
\begin{itemize}
    \item 3'-TTTTTGACTTGGA-5'
    \item 3'-TCCAAGTC-5'
\end{itemize}

Simulations in the standalone model were performed by setting the parameter ``diff\_coeff'' to 2.5, implying a diffusion coefficient of single particles of 2.5 simulation units. 

Simulation in LAMMPS were performed by setting the inertial timescale $\tau_{in}=1/\gamma$=2.5, which is controlled via the ``damp'' parameter in the Langevin thermostat. This corresponds to a diffusion coefficient $D=k_B T / m \gamma = 0.07918$ in LAMMPS simulation units.

    \begin{table}[]
\centering
\begin{tabular}{| >{\centering\arraybackslash} p{2cm}| >{\centering\arraybackslash} p{2cm}| >{\centering\arraybackslash} p{1.9cm}|
>{\centering\arraybackslash} p{1.9cm}| } 
 \hline
 \centering
$\Delta t$ & PE & KE\\

 \hline
 \hline
 0.001  & -1.265(2) &   0.2999(3) \\ 
 0.003  & -1.262(5) &   0.3004(11) \\ 
 0.005  &-1.263(3) &   0.3014(4) \\
 0.007 & -1.237(4) &   0.3067(6) \\
 VMMC & -1.2658(5) &   - \\

  \hline
\end{tabular}
\caption{ Mean potential and kinetic energies per nucleotide for MD simulations in the standalone code. The results for the potential energy in VMMC simulations are included for comparison. Brackets give standard error on the mean of final digit(s).}
\label{table:mean_std_energies}
\end{table}

 \begin{table}[]
\centering
\begin{tabular}{| >{\centering\arraybackslash} p{2cm}| >{\centering\arraybackslash} p{2cm}| >{\centering\arraybackslash} p{1.9cm}|
>{\centering\arraybackslash} p{1.9cm}| } 
 \hline
 \centering
$\Delta t$ & PE & KE \\

 \hline
 \hline
 0.001  & -1.243(6) &   0.29997(6) \\ 
 0.003  & -1.234(3) &   0.29994(4)\\ 
 0.005  &-1.248(8) &   0.29991(5)\\
 0.01 & -1.249(4) &   0.29995(4) \\
 VMMC & -1.2658(5) &   -  \\

  \hline
\end{tabular}
\caption{ Mean potential and kinetic  energies per nucleotide for MD simulations using the Langevin thermostat in the LAMMPS code (``damp'' set to 2.5). The results for the potential energy in VMMC simulations (performed in the standalone code)  are included for comparison. Brackets give standard error on the mean of final digit(s).}
\label{table:mean_std_energies2}
\end{table}

In Tables \ref{table:mean_std_energies} and \ref{table:mean_std_energies2}, the VMMC results for the potential energy are assumed to be accurate (to within sampling noise) since, unlike the MD algorithms, VMMC does not require a limit to be taken before its stationary distribution converges on $p_{\rm eq}(x)$. For the kinetic energy, equipartition gives the correct value as 0.3 in oxDNA units ($kT$ is 0.1, and each nucleotide has six momentum degrees of freedom)

Overall, for the standalone integrator performance has substantially worsened by $\Delta t=0.007$. $\Delta t=0.005$ performs reasonably, but it should be noted that stability issues sometimes arise when that time step is used in GPU simulations, perhaps due to extra errors introduced by the use of mixed precision. Generally, we would recommend $\Delta t=0.003$ as safe for this (typical) choice of ``diff\_coeff''.

The Langevin integrator shows no noticeable degradation in performance over the range of timesteps. For this choice of  the parameter ``Damp'', simulations of up to  timestep of $\Delta t = 0.01$ are feasible. For larger values of ``damp'' (actually equating to weaker damping), smaller timesteps are necessary to prevent instability.

\bibliography{references.bib}
\newpage

\end{document}